\documentclass[
 reprint,
superscriptaddress,
 amsmath,amssymb,
prd,
twocolumn
]{revtex4-2}
\usepackage{amsfonts,amssymb,amsmath}
\usepackage[caption=false]{subfig} %default subfig option screws up captions
\usepackage{graphicx, hyperref, tabularx, dcolumn}
\usepackage{mathtools}

\usepackage{color}
\usepackage[dvipsnames]{xcolor}
\usepackage{soul}
\usepackage{tensor, comment}
\usepackage{enumitem}
\usepackage[export]{adjustbox}

\begin{document}

\title{General-Relativistic Gauge-Invariant Magnetic Helicity Transport:\\
Basic Formulation and Application to Neutron Star Mergers}
\author{Jiaxi Wu}
\affiliation{TAPIR, Mailcode 350-17, California Institute of Technology, Pasadena, CA 91125, USA}
%\affiliation{Department of Physics, California Institute of Technology, Pasadena, CA 91125, USA}
\author{Elias R. Most}
\affiliation{TAPIR, Mailcode 350-17, California Institute of Technology, Pasadena, CA 91125, USA}
\affiliation{Walter Burke Institute for Theoretical Physics, California Institute of Technology, Pasadena, CA 91125, USA}
%\affiliation{Department of Physics, California Institute of Technology, Pasadena, CA 91125, USA}
%\date{\today}

\begin{abstract}
Dynamo processes are ubiquitous in astrophysical systems. In relativistic astrophysical systems, such as accretion disks around black holes or neutron stars, they may critically affect the launching of winds and jets that can power electromagnetic emission. Dynamo processes are governed by several microscopic parameters, one of them being magnetic helicity.
As a conserved quantity in nonresistive plasmas, magnetic helicity is transported across the system. One important implication of helicity conservation is, that in the absence of helicity fluxes some mean-field dynamos can be quenched, potentially affecting the large-scale evolution of the magnetic field. 
One of the major challenges in computing magnetic helicity is the need to fix a meaningful electromagnetic gauge.
We here present a fully covariant formulation of magnetic helicity transport in general-relativistic plasmas based on the concept of relative helicity by Berger \& Field and Finn \& Antonsen. This formulation is separately invariant under gauge-transformation of the Maxwell and Einstein equations.
As an application of this new formalism we present the first analysis of magnetic helicity transport in the merger of two neutron stars. We demonstrate the presence of global helicity fluxes into the outer layers of the stellar merger remnant, which may impact subsequent large-scale dynamo amplification in these regions.
\end{abstract}

%\keywords{Magnetic helicity, Neutron star}

\maketitle

\section{Introduction}
Astrophysical dynamos are important for a variety of systems, including
stars \cite{annurev:/content/journals/10.1146/annurev-astro-081913-040012}, accretion disks \cite{1995A&A...298..934R}, planets \cite{RevModPhys.72.1081} and galaxies \cite{annurev:/content/journals/10.1146/annurev-astro-071221-052807}. Dynamos can operate on different scales, causing turbulent magnetic field amplification on small (microscopic scales) to mean-field processes in large scale shear flows \cite{brandenburg2012current,Schekochihin_2022}. In relativistic systems, turbulent dynamo amplification may
aid the launching of jets and outflows from neutron stars \cite{Kiuchi:2017zzg,Kiuchi:2023obe,Combi:2023yav,Most:2023sft,Aguilera-Miret:2023qih,Palenzuela:2021gdo} and
black hole accretion disks \cite{Liska:2018btr,Hogg:2018zon,Jacquemin-Ide:2023qrj}.

While the details of these individual processes are complicated and will depend on the specific dynamo mechanism operating, several of them including the $\alpha\Omega$-dynamo \cite{Gruzinov:1994zz} potentially active in neutron star mergers \cite{Kiuchi:2023obe}, are affected by a magnetic field invariant, {\it magnetic helicity}.

Magnetic helicity is a topological invariant of the plasma \cite{1999PPCF...41B.167B}. In simple terms, it quantifies the amount of magnetic field line linkage in a given volume. A high degree of helicity implies very intricately intertwined field geometries. While helicity can be dissipated by resistivity, in nonresistive plasmas (such as in many relativistic systems) it is a globally conserved quantity. In other words, any magnetic field amplification or rearrangement will be subject to the constraints imposed by the initial helicity content in a given volume. 
Prominent examples of this concern the relaxation of a given field into its lowest energy configuration \cite{1974PhRvL..33.1139T}, which can affect large scale structures of the field.
In addition, Refs. \cite{Gruzinov:1994zz,1995ApJ...449..739B} (see also Ref. \cite{Blackman:2014kxa}) have argued that some dynamo actions may be quenched in the presence of finite helicity, such that the field may not be able to rearrange itself into large scale structures. This effect of $\alpha-$quenching can particularly affect $\alpha\Omega$-dynamos, though other dynamo models \cite{Squire:2015jma} potentially active in accretion disks \cite{Jacquemin-Ide:2023qrj} are intrinsically non-helical.\\

A potential way to circumvent the limitations of a fixed helicity inside a given volume are helicity fluxes, which allow helicity to be transported in and out of this volume.
Since many of the (relativistic) systems, we are interested in, feature complicated large scale flow structures, it is conceivable that such a situation may not be uncommon. 
With a wealth of numerical simulations available nowadays, this question can in principle be answered.

In practice, a major obstacle in computing helicity fluxes is the gauge dependence of any local helicity measurement \cite{Berger_Field_1984}. Put differently, while the global value of the helicity over a volume enclosing all magnetic field lines of the system  is well defined and gauge invariant, the value of its contributions throughout the volume is gauge dependent.

Addressing this problem is not straightforward. One possibility is to relax the helicity definition to mean the linkage with respect to a single field line, which can be more easily computed \cite{2019A&A...624A..51M}.
Previous numerical studies have commonly considered a fixed gauge, which may be appropriate given special geometries of the problem \cite{2009MNRAS.398.1414B}. For a general flow, as we plan to consider, such an approach may not always be feasible.
Apart from uncertainties in quantifying transport,  helicity conservation may also be affected by artificial dissipation of the numerical scheme \cite{Zenati:2021vrr}.
%ermcom{Include these somwhere: \cite{Blackman_2014,Simon_2011,Ebrahimi_2014}}

Leveraging advances in solar dynamo physics \cite{Schuck_2019}, in this work we formulate a general-relativistic version of gauge-invariant magnetic helicity transport in arbitrary spacetimes. This formulation relies on a clean separation of field lines enclosed in a subvolume from those leaving this volume in a formulation first developed by Berger \& Field \cite{Berger_Field_1984} and Finn \& Antonsen \cite{Finn_Antonsen_1985}. 

Our paper is structured as follows: In Sec. \ref{sec:picture}, we present a summary of the general idea of helicity and its transport, which we then mathematically formulate in Sec. \ref{sec:methods}. In Sec. \ref{sec:neutronstar}, we demonstrate how this formulation can be applied to a neutron star merger simulation.
Throughout this work, we use geometric units, $G=c=1$, and Heaviside-Lorentz units for the electromagnetic fields.
We further use Greek indices $(\mu,\nu,\alpha,\dots)$ to indicate indices running from $0$ to $3$, and Latin indices $(i,j,k,\dots)$ to represent spatial indices from $1$ to $3$.

\section{Basic picture}\label{sec:picture}

Magnetic helicity is a measurement of magnetic flux tube linkage, including terms of ``self-helicity" and ``mutual helicity".
The self-helicity quantifies how magnetic field lines are twisted and writhed with respect to itself, while the mutual helicity accounts for crossings of different field lines \cite{Moffatt_14}. 
Magnetic helicity can be defined in terms of a helicity density,
\begin{align}
    \mathcal{H} = \boldsymbol{A}\cdot\boldsymbol{B}\,,
\end{align}
where $\boldsymbol{A}$ is the magnetic vector potential, and $\boldsymbol{B}=\boldsymbol{\nabla}\times\boldsymbol{A}$ is the magnetic field.
While $\mathcal{H}$ can be computed everywhere in the domain, due to the gauge degree of freedom of the magnetic vector potential, $\boldsymbol{A}\rightarrow \boldsymbol{A}+\boldsymbol{\nabla}\Lambda$, the local helicity density $\mathcal{H}$ is fundamentally not gauge invariant.

At the same time, the global helicity,
\begin{equation}
H=\int_V{\text{d}^3x\,\mathcal{H}}=\int_V{\text{d}^3x\,\boldsymbol{A}\cdot\boldsymbol{B}}\,,\label{eq:3d_hel_def}
\end{equation}
is gauge invariant, provided that the volume $V$ encloses all magnetic field lines, i.e., 
\begin{equation}
H'=H+\int_V{\text{d}^3x\,\boldsymbol{\nabla}\Lambda\cdot\boldsymbol{B}}=H-\oint_{\partial V}{\text{d}S\,{\boldsymbol{s}}\cdot\Lambda\boldsymbol{B}},
\end{equation}
where the last term vanishes if $B$ is fully enclosed in the volume $V$. Here ${\boldsymbol{s}}$ is the unit normal pointing inward to the boundary $\partial V$.

However, in realistic situations one might be interested in a more local measure, i.e., the amount of helicity in the vicinity of a black hole rather than of the entire accretion disk. The helicity concept above can be generalized to the case, where one separates magnetic field lines entirely contained inside a volume $V$ from those leaving it through the boundary. Since from a measurement in $V$ alone one cannot say anything about these ``open" field lines, asking about linkage numbers for magnetic field lines entirely contained in a volume is meaningful within the helicity framework introduced above.

\begin{figure}
    \centering
    \includegraphics[width=0.95\linewidth]{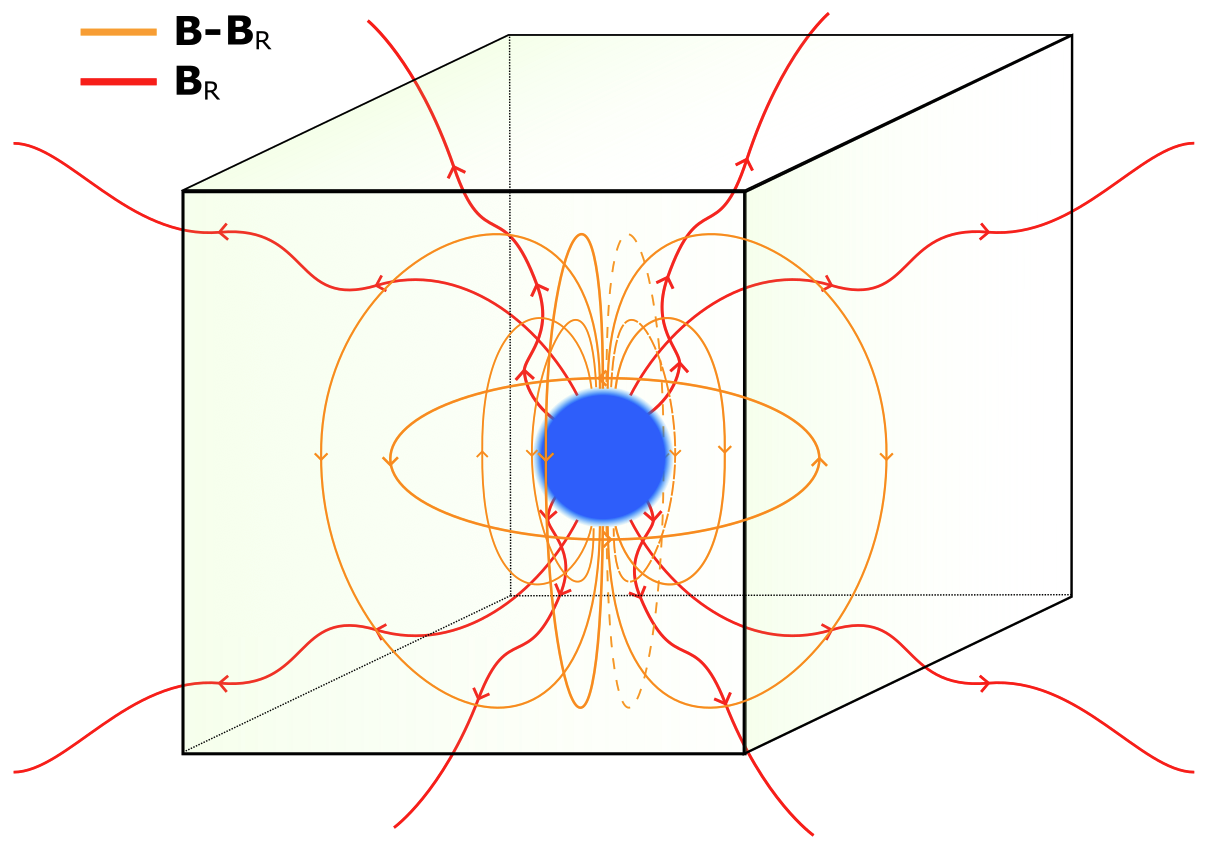}
    \caption{A sketch showing the concept of reference fields. Magnetic fields shown as orange field lines are completely confined within the volume boundary, and thus can be used to calculate gauge-independent relative helicity. In contrast, red field lines are crossing the boundary and can be subtracted out by introducing the reference field $\boldsymbol{B}_R$ corresponding to these field lines.}
    \label{fig:rel_hel_cartoon}
\end{figure}

This concept of {\it relative helicity} goes back to Berger \& Field \cite{Berger_Field_1984} and Finn \& Antonsen \cite{Finn_Antonsen_1985}. Schematically, the situation we are after is shown in Fig. \ref{fig:rel_hel_cartoon}. Here we can see the background field $\boldsymbol{B}_R$, which consists solely of field lines leaving the volume $V$ in question, separated from those entirely contained in it $\boldsymbol{B}-\boldsymbol{B}_R$. 
Following Ref. \cite{Berger_Field_1984}, the background field can be constructed as a potential flow, satisfying
\begin{align}
    \boldsymbol{B}_R &= -\boldsymbol{\nabla}\psi_R\,,\\
    \boldsymbol{\nabla}^2\psi_R &= 0\,,
\end{align}
such that,
\begin{equation}\label{eqn:BR_boundary}
    \left. {\boldsymbol{s}} \cdot \boldsymbol{B}\right|_{\partial V} = \left. {\boldsymbol{s}} \cdot \boldsymbol{B}_R \right|_{\partial V}= - \left. \frac{\partial \psi_R}{\partial s}\right|_{\partial V}\,,
\end{equation}
which ensures that the reference fields (red lines in Fig.~\ref{fig:rel_hel_cartoon}) match the outgoing magnetic field lines on the boundary $\partial V$ of the volume $V$.
Given the background magnetic field, one can uniquely define an associated magnetic vector potential $\boldsymbol{A}_R$, such that $\boldsymbol{B}_R = \boldsymbol{\nabla}\times \boldsymbol{A}_R$ and $\boldsymbol{\nabla}\cdot \boldsymbol{A}_R =0$.
Equipped with this split into background and foreground fields, the relative helicity density is defined as \cite{Berger_Field_1984,Finn_Antonsen_1985},
\begin{equation}
    h=(\boldsymbol{A}+\boldsymbol{A}_R)\cdot(\boldsymbol{B}-\boldsymbol{B}_R)\,.
\end{equation}

Following Refs. \cite{Berger_Field_1984,Finn_Antonsen_1985}, we can express the relative helicity as,
\begin{equation}\label{eq:3d_rel_hel_def}
    H= \int_V{\text{d}^3x\,h}=\int_V{\text{d}^3x\,(\boldsymbol{A}+\boldsymbol{A}_R)\cdot(\boldsymbol{B}-\boldsymbol{B}_R)}.
\end{equation}

It can be shown that Eq. \eqref{eq:3d_rel_hel_def} defined in this way is gauge-invariant and retains its usual meaning \cite{Berger_Field_1984,Finn_Antonsen_1985}.

Computing this term in practice is rather involved. First one needs to compute the background field, which will be the solution to a three-dimensional elliptic equation with the boundary conditions given by Eq. \eqref{eqn:BR_boundary}.
This is rather unwieldy in practice, especially if one wanted to do it for many subdomains of a numerical simulation. Relative magnetic helicity is also not additive, meaning that it has to be recomputed separately for every volume $V$ under considerations \cite{2020A&A...643A..26V}.

Computing the full relative helicity integral \eqref{eq:3d_rel_hel_def} is, however, not the 
only way to track helicity. It can be shown that the original helicity expression gives rise to a conservation law,
\begin{align}\label{eqn:helicity_transport}
    \frac{\partial \mathcal{H}}{\partial t} + \boldsymbol{\nabla}\cdot\left(\boldsymbol{E}\times \boldsymbol{A} - \Phi \boldsymbol{B}  \right) = - 2 \boldsymbol{E}\cdot \boldsymbol{B}\,,
\end{align}
where $\Phi$ is the scalar potential, and the RHS vanishes for a nonresistive plasma, for which $\boldsymbol{E}=-\boldsymbol{v}\times\boldsymbol{B}$. However, for a helical dynamo, e.g., an $\alpha$-term, $\boldsymbol{E}_{\rm dyn} \sim \kappa \boldsymbol{B}$, helicity production on the RHS has to (on average) be balanced by a net helicity flux to ensure global conservation. While the above transport equation is not locally gauge invariant, we can take a similar approach and re-formulate the relative helicity as a transport problem (relative to a given volume $V$) \cite{Schuck_2019},
\begin{eqnarray}\label{eqn:schuck_transport}
    \frac{\partial H}{\partial t}&=&-2\int_{{V}}{\text{d}^3x\,\boldsymbol{E}\cdot\boldsymbol{B}}\notag\\
    &&-\oint_{\partial V}{\text{d}S\, {\boldsymbol{s}}\cdot[(\boldsymbol{A}+\boldsymbol{A}_R)\times(\boldsymbol{E}-\boldsymbol{E}_R)]}.
\end{eqnarray}
Due to the inclusion of the background field, this expression fundamentally maintains gauge invariance (see Appendix \ref{sec:gauge_invariance}).

While the above expression does at first glance seemingly not simplify our problem, it turns out that after substantial algebraic manipulation in the case of an ideal (nonresistive) plasma one can recast this expression as \cite{Schuck_2019},
\begin{equation}
    \frac{\partial H_S}{\partial t}=2\oint_{\partial V}{\text{d}S\,\zeta B_n},\label{eq:schuck_surface_transport}
\end{equation}
where $\zeta$ is the lamellar part (i.e. the part with zero normal component of the curl of the surface gradient) of the electric field, which can be calculated by solving a surface elliptic partial differential equation on the boundary $\partial V$ of the enclosing volume,
\begin{equation}
    \nabla_S^2\zeta=-\boldsymbol{\nabla}_S\cdot\boldsymbol{E}_S.\label{eq:schuck_surface_pde}
\end{equation}
We denote the induced two-dimensional operators and fields on the boundary with a subscript $S$.  In particular, for a spherical volume $\zeta$ is uniquely determined by the solution of Eq.~\eqref{eq:schuck_surface_pde}. 
Remarkably, this equation only depends on the ideal electric field, not on the vector potential, $\boldsymbol{A}$, making it suitable also for most numerical codes that only evolve $\boldsymbol{B}$ \cite{Font:2008fka}. It should not come as a surprise that this expression must involve a surface elliptic equation, since the reference magnetic field approach in essence counts the number of open magnetic field lines, which is a global problem on the boundary.

In the remainder of this work, we will generalize this approach \cite{Schuck_2019} to general-relativity in arbitrary spacetimes. We will show that when carefully defining the relevant electric and magnetic fields within a 3+1 decomposition of spacetime, we can retain the the simplicity character of Eqs. \eqref{eq:schuck_surface_transport} and \eqref{eq:schuck_surface_pde}.

\section{General-relativistic helicity transport}\label{sec:methods}

In this section, we want to derive a general-relativistic expression for gauge-invariant magnetic helicity transport.
Before we do so, we will briefly review the covariant form of the Maxwell equations, as well as decomposition of spacetime useful for our purposes.

\subsection{General-relativistic electrodynamics}\label{sec:GREM}

Covariant electrodynamics can be formulated in terms of a covariant vector potential $\mathcal{A}_\mu$, which gives rise to a field strength tensor,
\begin{align}
    F^{\mu\nu} &= \nabla^\mu \mathcal{A}^\nu - \nabla^\nu \mathcal{A}^\mu\,, 
\end{align}
where $\nabla_\mu$ denotes the covariant derivative relative to the four-dimensional spacetime metric $g_{\mu\nu}$.
It will turn out to be convenient to further introduce the dual of the field strength tensor,
\begin{eqnarray}
    &{~\!^{\ast}\!F}^{\alpha\beta}=\frac{1}{2}\varepsilon^{\alpha\beta\mu\nu}F_{\mu\nu},\\
\end{eqnarray}
where $\varepsilon^{\alpha\beta\mu\nu}=-\frac{1}{\sqrt{-g}}\epsilon^{\alpha\beta\mu\nu}$ is the four-dimensional Levi-Civita tensor, and $\epsilon^{\alpha\beta\mu\nu}$ is the permutation symbol. The corresponding covariant Levi-Civita tensor is $\varepsilon_{\alpha\beta\mu\nu}=\sqrt{-g}\epsilon_{\alpha\beta\mu\nu}$.

The evolution equations for the field strength tensor are given by the covariant form of the Maxwell equations,
\begin{align}\label{eqn:Maxwell}
    \nabla_\beta F^{\alpha\beta}&=\mathcal{J}^\alpha\,,\\
    \nabla_\beta \prescript{\ast}{}{F}^{\alpha\beta}&=0\,,
\end{align}
where $\mathcal{J}^\alpha$ is the electric current.

We can now define a helicity current \cite{Bekenstein87},
\begin{align}
    \mathcal{H}^\mu = \mathcal{A}_\nu {~\!^{\ast}\!F}^{\mu\nu}\,.\label{eq:4d_hel_def}
\end{align}
Using the Maxwell equations \eqref{eqn:Maxwell}, it follows that,
\begin{align}\label{eqn:cov_helicity_transport}
    \nabla_\mu \mathcal{H}^\mu = {~\!^{\ast}\!F}^{\alpha\beta} F_{\alpha\beta}\,,
\end{align}
which is the covariant generalization of the helicity transport equation \eqref{eqn:helicity_transport}.

The energy evolution of the electromagnetic field can be associated with an energy momentum tensor,
\begin{align}
    T_{\rm EM}^{\mu\nu} = F^{\mu\alpha} F^\nu_\alpha - \frac{1}{4}g^{\mu\nu} F_{\alpha \beta} F^{\alpha\beta}\,,
\end{align}
which satisfies \cite{Baumgarte:2002vv},
\begin{align}
    \nabla_\mu T^{\nu\mu}_{\rm EM} = - F^{\mu\nu} \mathcal{J}_\mu\,.
\end{align}

\subsubsection{3+1 decomposition of spacetime}
In order to make a connection to the formulation in terms of electric and magnetic fields discussed in Sec. \ref{sec:picture}, we need to introduce a normal observer, $ n_\mu= (-\alpha,0,0,0)$, whose worldline trajectory will provide a time direction. Here $\alpha$ is the lapse function.
We do so using the conventional 3+1 decomposition of spacetime \cite{Arnowitt:1962hi}, where the metric is expressed as,
\begin{equation}
    g_{\mu\nu}\text{d}x^\mu\text{d}x^\nu=-\left(\alpha^2 -\beta_i \beta^i\right)\text{d}t^2+ 2\beta_i\text{d}x^i\text{d}t + \gamma_{ij}\text{d}x^i\text{d}x^j\,,
\end{equation}
with spatial metric $\gamma_{\mu\nu}=g_{\mu\nu}+n_\mu n_\nu$, $n^{\nu}=(1/\alpha,-\beta^i/\alpha)$, and $\beta^i$ being the coordinate shift. 

Within this coordinate choice, we can split the covariant vector potential $\mathcal{A}_\mu$ and electric current $\mathcal{J}_\mu$ into, 
\begin{align}
    \mathcal{A}_\mu &=\Phi n_\mu+A_\mu\,,\\
    \mathcal{J}_\mu &=q n_\mu+J_\mu\,,
\end{align}
where $\Phi$ is the scalar potential, $A_\mu$ the vector potential, $q$ the electric charge density, and $J_\mu$ the electric current on the three dimensional hypersurface induced by $n_\mu$.

Using this split, we can introduce electric, $E^\mu$, and magnetic, $B^\mu$, fields via \cite{Baumgarte:2002vv},
\begin{align}
    E^\mu &= n_\nu F^{\mu\nu} = \alpha F^{0\mu}\,,\\
    B^\mu &= n_\nu {~\!^{\ast}\!F}^{\mu\nu} = \alpha {~\!^{\ast}\!F}^{0\mu}\,.
\end{align}
One can show that,
\begin{align}\label{eqn:BfromA}
    B^i = \varepsilon^{ijk} \partial_j A_k\,,
\end{align}
 where $\varepsilon^{\mu\nu\kappa}= n_\alpha \varepsilon^{\mu\nu\kappa\alpha}$ is the three-dimensional Levi-Civita tensor. This 
establishes consistency between the 3+1 formulation of electrodynamics and those in flat spacetimes.

We can similarly decompose the field strength tensor into \cite{Baumgarte:2002vv},
\begin{align}
    F^{\alpha\beta}&=n^\alpha E^\beta-n^\beta E^\alpha-\varepsilon^{\alpha\beta\mu\nu}B_\mu n_\nu\,,
     \label{eq:maxwell_tensor}\\
    \prescript{\ast}{}{F}^{\alpha\beta}&=-n^\alpha B^\beta +B^\alpha n^\beta -\varepsilon^{\alpha\beta\mu\nu}E_\mu n_\nu\,.
    \label{eq:faraday_tensor}
\end{align}

Within the 3+1 split, the Maxwell equations take the familiar form,
\begin{eqnarray}
    &\mathcal{D}_i E^i=q,\\
    &\mathcal{D}_i B^i=0, \label{eqn:divB}\\
    &\partial_t(\sqrt{\gamma}E^i)-\epsilon^{ijk}\partial_j \tilde{B}_k=-\sqrt{\gamma} J^i,\\
    &\partial_t(\sqrt{\gamma}B^i)+\epsilon^{ijk}\partial_j D_k=0.
\end{eqnarray}
Here $\mathcal{D}_i x^i=\frac{1}{\sqrt{\gamma}}\partial_i(\sqrt{\gamma}x^i)$ is the three-dimensional covariant derivative with respect to $\gamma_{ij}$. 
Finally, we have introduced effective expressions for the electric and magnetic fields that enter the flux terms in the above expressions,
\begin{eqnarray}
    &D^i=\alpha E^i+\varepsilon^{ijk}\beta_jB_k,\label{eq:relation_DE}\\
    &\tilde{B}^i=\alpha B^i-\varepsilon^{ijk}\beta_jE_k.\label{eq:relation_HB}
\end{eqnarray}
When using such definitions, it can be shown that the Maxwell equations take a form closest to their flat spacetime counterparts \cite{Komissarov:2004ms}.

\subsubsection{Surface Helicity Transport}
With the formalism for covariant electrodynamics in place, we are now in a position to formulate relative helicity transport within the 3+1 language. 
Following our discussion in Sec. \ref{sec:picture}, we need to define 
a reference magnetic field, $B^i_R = -\mathcal{D}^i \psi_R$, for a suitable scalar potential $\psi_R$, where
\begin{eqnarray}
   -\mathcal{D}_i \mathcal{D}^i \psi_R= \mathcal{D}_i B_R^i=0,&\qquad &s_iB^i|_S=s_iB_R^i|_{\partial V},\label{eq:ref_fd_bd1}\\
    \mathcal{D}_iA_R^i=0,&\qquad &s_iA_R^i|_{\partial V}=0\,,\label{eq:ref_fd_bd2}
\end{eqnarray}
with a corresponding vector potential, $\mathcal{A}^\mu_R = A^\mu_R$, such that $B^i_R = \varepsilon^{ijk} \partial_j \left(A_{R}\right)_k$, and vanishing scalar potential $\Phi_R=0$.
The presence of relativity does almost not affect these equations, except through the necessity of using three-dimensional covariant derivatives, $ \mathcal{D}_i$ in order to preserve the constraint condition \eqref{eqn:divB}. 

In relativity, we cannot express the helicity density in terms of the magnetic field, but need the full field strength tensor. We therefore define the corresponding field strength tensor of the background field as
\begin{align}
    F^{\mu\nu}_R = \nabla^\mu \mathcal{A}^\nu_R - \nabla^\nu \mathcal{A}^\mu_R \,.
\end{align}
Using these definitions, we propose to generalize the helicity density \eqref{eq:3d_rel_hel_def} to a relativistic relative helicity current, $h^\mu$ in the following way
\begin{equation}
    h^\mu=(\mathcal{A}_\nu+\mathcal{A}_{R\nu})(\prescript{\ast}{}{F}^{\mu\nu} - \prescript{\ast}{}{F_R}^{\mu\nu})\,.\label{eq:4d_rel_hel}
\end{equation}

For easier comparison with their non-relativistic expressions, we now expand this expression into the normal electric and magnetic fields,
\begin{eqnarray}
    h^0&=&\frac{1}{\alpha}(A_j+A_{Rj})(B^j-B_R^j)  ,\label{eq:comp_form_hel_t}\\
    h^i&=& \Phi (B^i-B_R^i) -\frac{1}{\alpha}\varepsilon^{ijk}(A_j+A_{Rj})(D_k-D_{Rk}).
%    h^i&=&\left[\phi+\phi_R+\frac{1}{\alpha}(A_j+A_{Rj})\beta^j\right](B^i-B_R^i)\notag\\
%    &&-\frac{1}{\alpha}\beta^i(A_j+A_{Rj})(B^j-B_R^j)\notag\\
%    &&-\varepsilon^{ijk}(A_j+A_{Rj})(E_k-E_{Rk}). 
    \label{eq:comp_form_hel_spc}
\end{eqnarray}

We point out that although $h^\mu$ in Eq.~\eqref{eq:4d_rel_hel} may appear to be a regular vector field, it is only well defined relative to a given volume $V$, on the boundary $\partial V$ of which we have calibrated the background field. 

Next, we want to compute the corresponding transport equation for the relative helicity current $h^\mu$. We now compute the four-covariant divergence of $h^\mu$,
\begin{equation}\label{eqn:3p1_transport}
    \partial_\mu (\alpha\sqrt{\gamma}h^\mu)=-2 \sqrt{\gamma} (D_i B^i-D_{Ri} B_R^i)\,.
\end{equation}
which conceptually takes the same form as \eqref{eqn:cov_helicity_transport}.

Based on this equation, we can now generalize the definition of relative helicity to the relativistic context,
\begin{align}
    \mathcal{H} &= \alpha \sqrt{\gamma} h^0\,,\\
    H &= \int_{V}\! {\rm d}^3 x\, \mathcal{H} = \int_{V}\! {\rm d}^3 x\, \sqrt{\gamma} (A_j+A_{Rj})(B^j-B_R^j)\,.
\end{align}

Combining this definition with the transport equation \eqref{eqn:3p1_transport}, we find
\begin{eqnarray}\label{eqn:cov_transport_master}
     \partial_t H&=&
    -2\int_V{\text{d}^3x\,\sqrt{\gamma}(D_iB^i-D_{Ri}B_R^i)}\notag\\
    &&-\oint_{\partial V}{\text{d}Ss_i\sqrt{\gamma}\varepsilon^{ijk}(A_j+A_{Rj})(D_k-D_{Rk})}\,,\label{eq:hel_evol_DB}
\end{eqnarray}
which is a direct generalization of the main transport equation \eqref{eqn:schuck_transport} of Ref. \cite{Schuck_2019}.

In the last stage of obtaining the final form of the transport equation, we want to convert Eq. \eqref{eqn:cov_transport_master} into an equation resembling the surface Laplace problem \eqref{eq:schuck_surface_transport}. The main obstacle is to remove the direct appearance of the vector potential.

Following Ref. \cite{Schuck_2019}, we perform two decompositions with respect to the potential field $D_{Ri}$. Below we only outline the decompositions and corresponding boundary conditions, with details given in Appendix \ref{section:Append_decomp}. 
We first split the electric field into a solenoidal part, $\Sigma_R^i$, and an irrotational part, $\mathcal{D}^i \Lambda_R$, using a Helmholtz decomposition \cite{Schuck_2019}, i.e.,
\begin{equation}
D_R^i=\Sigma_R^i+\mathcal{D}^i\Lambda_R,\label{eq:first_decomp}
\end{equation}
with boundary constraints listed below
\begin{eqnarray}
    \mathcal{D}_i\Sigma_R^i=0,&\qquad & {s}_i\Sigma_R^i|_{\partial V}=0,\label{eq:helmholtz_decomp_bd1}\\
\mathcal{D}_i\mathcal{D}^i\Lambda_R=\mathcal{D}_iD_R^i,&\qquad &\Lambda_R|_{\partial V}=\text{constant},\label{eq:helmholtz_decomp_bd2}
\end{eqnarray}
so that $\boldsymbol{\Sigma}_R$ is uniquely determined and the irrotational part has no contribution to the helicity transport. In this case, Eq.~\eqref{eq:hel_evol_DB} can be written as
\begin{eqnarray}
    \partial_t H &=&-2\int_V{\text{d}^3x\,\sqrt{\gamma}D_iB^i}\notag\\
    &-&\oint_{\partial V}{\text{d}S\,s_i\sqrt{\gamma}\varepsilon^{ijk}(A_j+A_{Rj})(D_k-\Sigma_{Rk}}).\label{eq:hel_evol_1st_simp}
\end{eqnarray}
This expression contains two parts, $\partial_t H =\partial_t H_V+\partial_t H_S$, namely the volume change in helicity due to dissipation
\begin{align}
    \partial_t H_V = -2\int_V{\text{d}^3x\,\sqrt{\gamma}D_iB^i} =-2\int_V{\text{d}^3x\,\alpha\sqrt{\gamma}E_iB^i} \,,
\end{align}
which vanishes for an ideal plasma, as well as the surface transport term,
\begin{align}
    \partial_t H_S = -\oint_{\partial V}{\text{d}S\,s_i\sqrt{\gamma}\varepsilon^{ijk}(A_j+A_{Rj})(D_k-\Sigma_{Rk}})\,.
\end{align}
In turbulent systems, helicity will be produced on resistive scales due to the volume term, whereas on larger scales transport will take over. The volume term is gauge invariant and can be linked to magnetic current helicity \cite{2019ApJ...884...55R}. The transport part is gauge dependent, and it is the latter we want to analyze here.

Following Ref. \cite{Schuck_2019}, we then decompose $\boldsymbol{D}$ and $\boldsymbol{\Sigma}_R$ using the Helmholtz-Hodge theorem into a normal part $\tau \hat{\boldsymbol{s}}$, a solenoidal part $\hat{\boldsymbol{s}}\times\partial_{S}\frac{\partial\chi}{\partial t}$, and a lamellar component (i.e. have vanishing normal component of surface curl) $\partial_{S}\zeta$ with surface covariant derivative operator defined as $\partial_{Si}=\partial_i-s_i s^j \partial_j$ \cite{Schuck_2019},
\begin{eqnarray}
    &D_{i}=\tau s_i-\varepsilon_{ijk}s^j\partial_S^k\frac{\partial\chi}{\partial t}-\partial_{Si}\zeta,\label{eq:helmholtz-hodge_decomp}\\
    &\Sigma_{Ri}=\tau_R s_i-\varepsilon_{ijk}s^j\partial_S^k\frac{\partial\chi_R}{\partial t}-\partial_{Si}\zeta_R.\label{eq:helmholtz-hodge_decomp_ref}
\end{eqnarray}
The three components for $\boldsymbol{D}$ (and similarly for $\boldsymbol{\Sigma}_R$) can be calculated from \cite{Schuck_2019},
\begin{eqnarray}
    &\tau = s^iD_i,\label{eq:norm_comp}\\
    &\partial_S^2\partial_t\chi=-s_i\varepsilon^{ijk}\partial_{Sj}(D_k-\tau s_k),\label{eq:PDE_of_chi}\\
    &\partial_S^2\zeta=-\partial_{Si}(D^i-\tau s^i).\label{eq:PDE_of_zeta}
\end{eqnarray}

Substituting these expressions into Eq.~\eqref{eq:hel_evol_1st_simp} and after some algebra (see Appendix \ref{app:helmholtz}), we find
\begin{eqnarray}
    \partial_t H_S    &=&\oint_{\partial V}{\text{d}S\,s_i\sqrt{\gamma}\varepsilon^{ijk}(A_j+A_{Rj})\partial_k(\zeta-\zeta_R)}\notag\\
    &=&2\oint_{\partial V}{\text{d}S\,\sqrt{\gamma}(\zeta-\zeta_R)B_n}\,,\label{eq:hel_evol_2nd_simp}
\end{eqnarray}
where $B_n = s_iB^i$ is the magnetic field component normal to the surface $\partial V$.
Refs. \cite{Schuck_2019} and \cite{Backus_1986} showed that for simple geometries (including spherical surfaces we consider in this work), $\zeta_R$ is a constant on the boundary. Thus, the term $\oint_{\partial V}{\text{d}S\,\sqrt{\gamma}B_n\zeta_R}$ vanishes due to solenoidal constraint on the magnetic field. The surface helicity transport is simply
\begin{equation}
    \partial_t H_S =2\oint_{\partial V}{\text{d}S\,\sqrt{\gamma}\zeta B_n},\label{eq:hel_evol_simp}
\end{equation}
where $\zeta$ can be solved by the surface Laplacian Eq.~\eqref{eq:PDE_of_zeta}.
This is the main equation for transport of relativistic gauge-invariant helicity through a given closed surface.

\subsection{Ideal magnetohydrodynamics limit}
\label{section:surface_laplacian}

We now consider the case of a nonresistive, ideal plasma with a fluid four-velocity $u^\mu$.
Within the 3+1 split, we can define an induced three-velocity, $v^i$, on the hypersurface, which satisfies
\begin{align}
    \tilde{u}^i \equiv \frac{u^i}{u^0} =  \alpha v^i - \beta^i\,,
\end{align}
where we call $\tilde{u}^i$ the advection velocity.
In ideal magnetohydrodynamics, the electric field is given as
\begin{align}
    E^i =& -\varepsilon^{ijk}v_j B_k\,,\\
    D^i =&-\varepsilon^{ijk}\Tilde{u}_jB_k\,. \label{eq:ohm'slaw}
\end{align}

We can further split the advection velocity into a part perpendicular to the magnetic field $\boldsymbol{\Tilde{u}}_\perp$, and a part parallel to the field $\boldsymbol{\Tilde{u}}_\parallel$. Then we can rewrite Eq.~\eqref{eq:ohm'slaw} into
\begin{equation}
    D^i= -\varepsilon^{ijk}\Tilde{u}_{\perp j}B_k \,,\label{eq:ohm'slaw_perp}
\end{equation}
where the perpendicular component can be calculated from
\begin{equation}
    \Tilde{u}_{\perp j}=\Tilde{u}_j-\frac{\Tilde{u}_iB^i}{B_kB^k}B_j\,, \label{eq:perp_v}
\end{equation}

We can further decompose the electric field into an emerging (em) term $D_{\rm em}^i$ (representing the transport of linked flux across the surface) and a shearing (sh) term $D_{\rm sh}^i$ (representing the twisting and tangling of footpoints by motions in the surface). These two terms depends on the normal and tangent components of the fluid advection velocity $\Tilde{u}_{\perp j}$, respectively. Therefore, we have
\begin{equation}
    D_{\text{em}}^i=-\varepsilon^{ijk}\Tilde{u}_{\perp_nj}(B_{nk}+B_{Sk})= -\varepsilon^{ijk}\Tilde{u}_{\perp_nj}B_{Sk}\,,\label{eq:emerging_elec}
\end{equation}
\begin{equation}
    D_{\text{sh}}^i=-\varepsilon^{ijk}\Tilde{u}_{\perp_Sj}(B_{nk}+B_{Sk})\,, \label{eq:shearing_elec}
\end{equation}
where $\Tilde{u}_{\perp j}$ and $B_k$ are separated into
\begin{equation}
    \Tilde{u}_{\perp_nj}=\Tilde{u}_{\perp i}s^i s_j\qquad \Tilde{u}_{\perp_Sj}=\Tilde{u}_{\perp j}-\Tilde{u}_{\perp_nj}\,,
\end{equation}
\begin{equation}
    B_{nk}=B_i s^i s_k\qquad B_{Sk}=B_k-B_{nk}\,,
\end{equation}
Combining Eqs.~\eqref{eq:emerging_elec} and \eqref{eq:shearing_elec} with Eq.~\eqref{eq:PDE_of_zeta}, we obtain the equations for the corresponding lamellar part $\zeta_{\rm em}$ and $\zeta_{\rm sh}$
\begin{align}
    \partial_S^2\zeta_{\text{em}}&=\partial_{Si}\varepsilon^{ijk}\Tilde{u}_{\perp_nj}B_{Sk} \,,\label{eq:laplace_em}\\
   \partial_S^2\zeta_{\text{sh}}&=\partial_{Si}\varepsilon^{ijk}\Tilde{u}_{\perp_Sj}B_{nk}\,,
   \label{eq:laplace_sh}
\end{align}
Note that $\varepsilon^{ijk}\Tilde{u}_{\perp_Sj}B_{Sk}$ in Eq.~\eqref{eq:shearing_elec} is discarded since that the surface divergence has no components along the unit normal direction. Combining Eqs.~\eqref{eq:laplace_em} and \eqref{eq:laplace_sh} into the surface transport Eq.~\eqref{eq:hel_evol_simp}, we obtain the final surface transport equation for magnetic helicity
\begin{equation}
    \partial_t H_S
    =2\oint_{\partial V}{\text{d}S\,\sqrt{\gamma}\left(\zeta_{\rm em} + \zeta_{\rm sh}\right) B_n} \,,\label{eq:surface_transport}
\end{equation}
and each term can be calculated separately in a gauge invariant way.

\section{General-Relativistic Helicity transport in numerical simulations}\label{sec:neutronstar}

Having derived a gauge-invariant formulation of relativistic relative helicity transport, we now want to demonstrate how these expressions can be used as a diagnostic tool in numerical simulations.
While the formulation we present is generic and can be applied to any simulation of turbulence, black hole accretion or other relativistic system, as a test problem we focus on neutron star mergers \cite{Baiotti:2016qnr,Radice:2020ddv}.
These feature complex (non-steady state) flow structures \cite{Kastaun:2016yaf}, small \cite{Kiuchi:2015sga,Palenzuela:2021gdo} and large-scale \cite{Kiuchi:2023obe,Most:2023sme} magnetic dynamo amplification, and present an ideal test bed to investigate global flows of magnetic helicity.

We begin by formulating the discrete problem of helicity transport on a coordinate sphere, before we apply it to a neutron star merger simulation.

\subsection{Helicity transport through spherical surfaces}
\label{section:discretization}
In simulations, it is convenient to extract data on coordinate spheres, either because the simulations are already using spherical coordinate, or because infrastructure readily exists to monitor winds and outflows on spheres. Furthermore, the transport equations we aim to solve require the existence of a Laplace operator, which is trivially fulfilled on such a smooth surface.

In spherical polar coordinates the surface divergence operator can be written as
\begin{equation}
    \partial_{Si}f^i=\frac{1}{r\sin{\theta}}\left[\frac{\partial}{\partial\theta}(\sin{\theta}f_\theta)+\frac{\partial f_\phi}{\partial\phi}\right]\,,
\end{equation}
whereas the Laplace operator takes the following form
\begin{equation}\label{eqn:laplace_surface}
    \partial_S^2\zeta=\frac{1}{r^2}\left[\frac{\partial}{\sin{\theta}\partial\theta}\left(\sin{\theta}\frac{\partial \zeta}{\partial\theta}\right)+\frac{1}{\sin^2{\theta}}\frac{\partial^2\zeta}{\partial\phi^2}\right]\,.
\end{equation}
Here $\theta$ is the longitudinal and $\phi$ is azimuthal coordinate.

We adopt a second-order finite difference method to discretize these surface operator in spherical polar coordinates.
Introducing grid coordinates $(i,j)$ on the numerical lattice, we can write
\begin{eqnarray}
    \partial_{Si}f^i&=&\frac{1}{r}\left(\frac{f_\theta^{i+1,j}-f_\theta^{i-1,j}}{2\Delta\theta}+\cot{\theta^{i,j}}f_\theta^{i,j}\right)\notag\\
    &&+\frac{1}{r\sin{\theta^{i,j}}}\frac{f_\phi^{i,j+1}-f_\phi^{i,j-1}}{2\Delta\phi}\,,\label{eq:div_discrete}
\end{eqnarray}
\begin{eqnarray}
    \partial_S^2\zeta&=&\frac{1}{r^2}\left(\frac{\zeta^{i+1,j}-2\zeta^{i,j}+\zeta^{i-1,j}}{\Delta\theta^2}+\cot{\theta^{i,j}}\frac{\zeta^{i+1,j}-\zeta^{i-1,j}}{2\Delta\theta}\right)\notag\\
    &&+\frac{1}{r^2\sin^2{\theta^{i,j}}}\frac{\zeta^{i,j+1}-2\zeta^{i,j}+\zeta^{i,j-1}}{\Delta\phi^2}\,,\label{eq:laplace_discrete}
\end{eqnarray}
where $\Delta \phi$ and $\Delta \theta$ are the discrete spacings of the numerical grid.

We then numerically solve Eqs. \eqref{eq:laplace_em} and \eqref{eq:laplace_sh}. For this first test, we use $236\times 116$ points, which makes a direct inversion of the resulting linear problem easily possible. In higher-resolution simulations with finer scale helicity structures, one would have to use higher angular resolutions. The helicity fluxes then straightforwardly follow from \eqref{eq:surface_transport}.

\begin{figure*}
    \centering
    \includegraphics[width=0.9\linewidth]{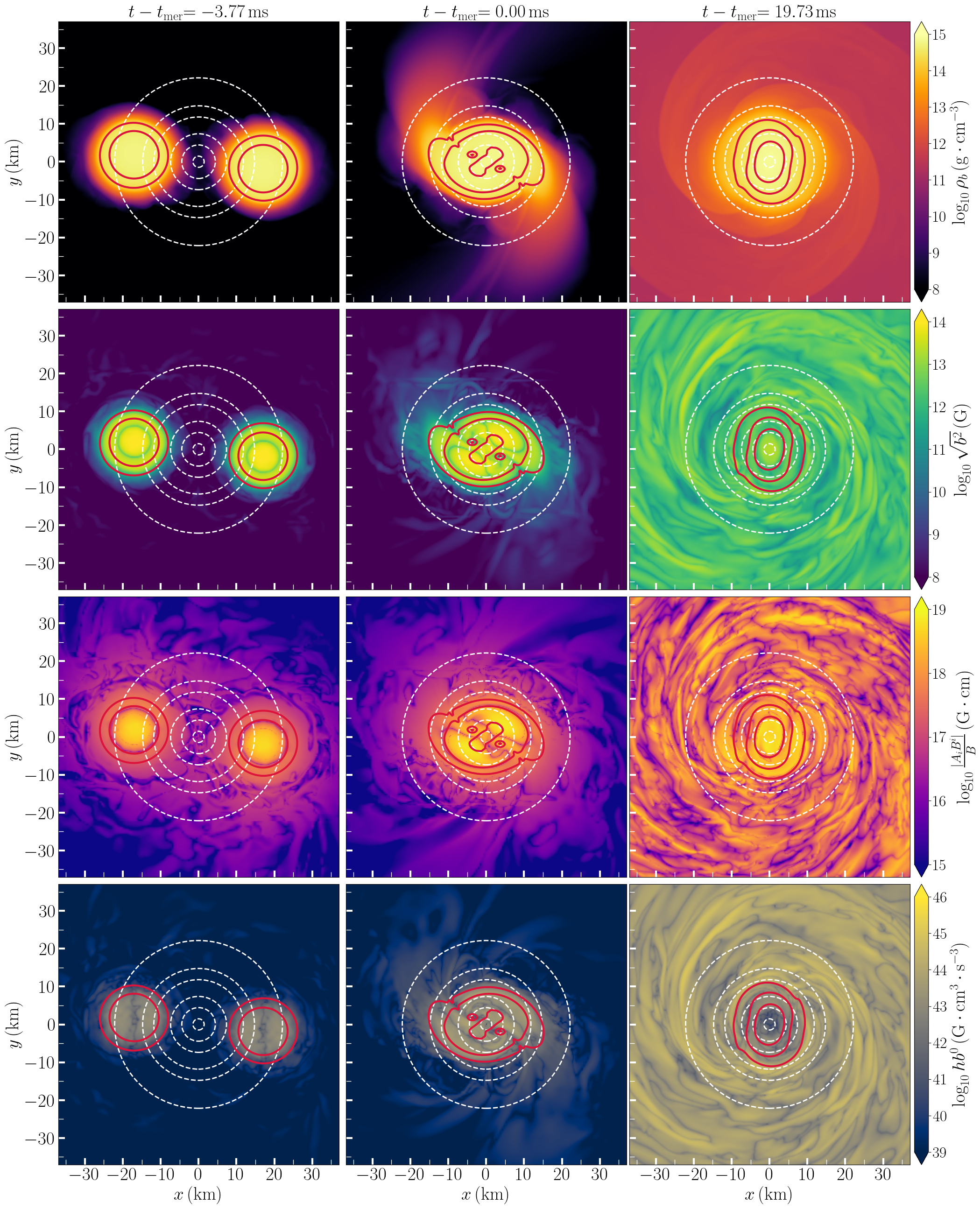}
    \caption{Evolution of the magnetic helicity density throughout an equal mass neutron star merger in the orbital plane. The four rows show the time evolution of rest mass density $\rho_b$, comoving magnetic field strength $\sqrt{b^2}$, normalized helicity density $A_iB^i/B$, and cross helicity $h b^0$, respectively. Dashed white circles mark the different spheres for which we compute the relative helicity. From outside in, red contour lines denote rest-mass density levels of $n_{\rm sat}, 2 n_{\rm sat}, 3 n_{\rm sat}$, respectively, where $n_{\rm sat}$ is the nuclear saturation density. All times, $t$, are stated relative to the time $t_{\rm mer}$ of merger.}
    \label{fig:2D_plot}
\end{figure*}

\begin{figure*}
    \centering
    \includegraphics[width=0.95\linewidth]{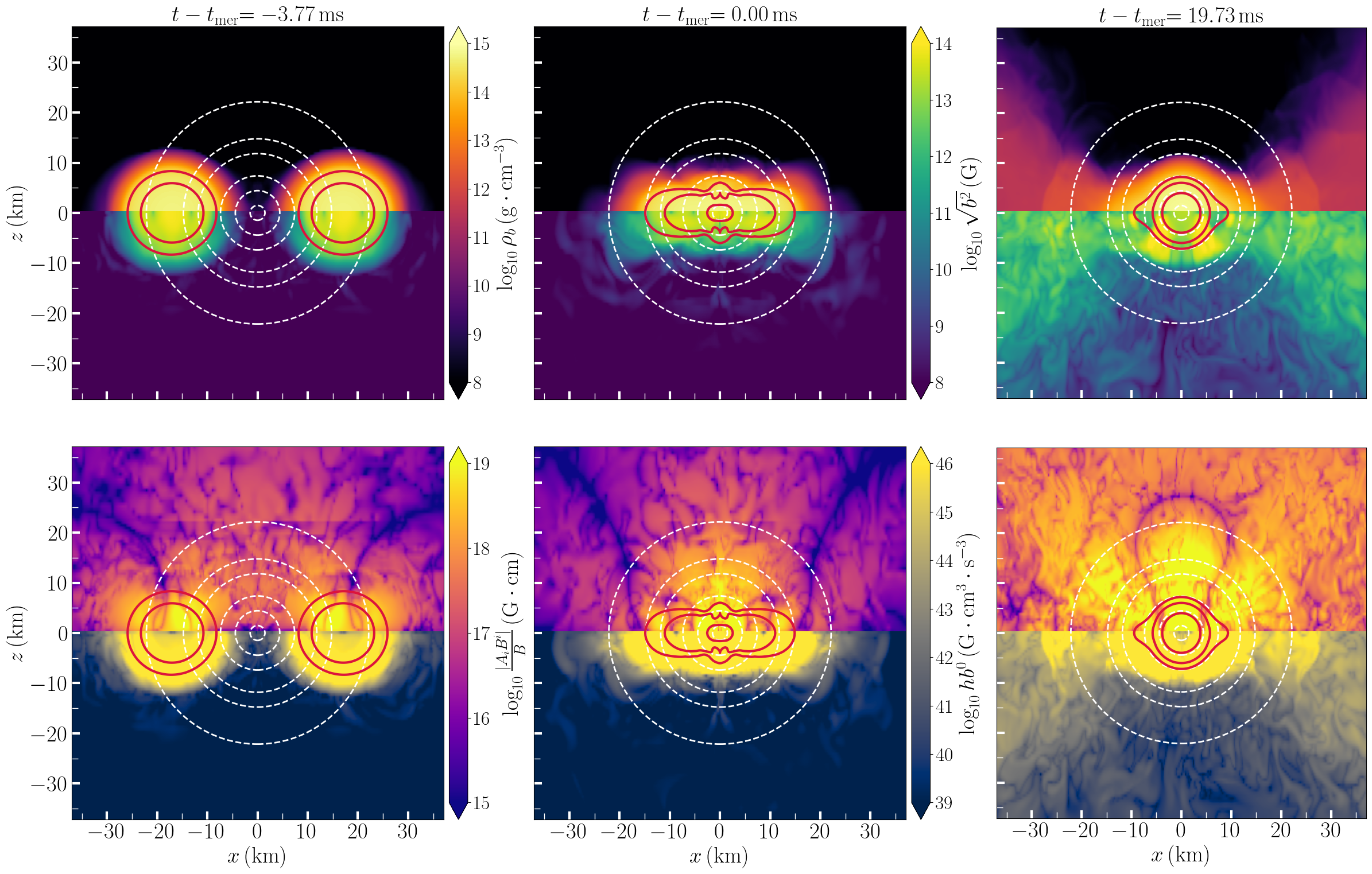}
    \caption{Same as Fig.~\ref{fig:2D_plot}, but in the meridional plane.}
    \label{fig:2D_plot_meridional}
\end{figure*}
\subsection{Helicity transport in neutron star mergers}

We here present a first demonstration of the helicity transport framework in a binary neutron star merger.
To this end, we have numerically evolved an equal mass binary neutron star system through merger using full numerical relativity to capture the spacetime and magnetohydrodynamics of the system. A detailed discussion of the numerical setup can be found in Appendix \ref{app:numerics}.

The magnetic field evolution in a neutron star merger has been studied
extensively (see,e.g., \cite{Baiotti:2016qnr} for a review). Rather than giving a full account of the merger and
post-merger dynamics, we only give a brief summary and defer to dedicated
studies in the literature \cite{Palenzuela:2015dqa,Kiuchi:2017zzg,Ciolfi:2019fie,Ruiz:2016rai}, since our main focus is on helicity transport.

Starting out with a mixed poloidal and toroidal field in each initial neutron star, we evolve the system through merger. Due to the disruption dynamics, the field is stretched leading to large toroidal fields in the remnant \cite{Ciolfi:2019fie}, which likely dominate over the pre-merger topology \cite{Aguilera-Miret:2021fre}. This is because strong turbulent dynamo amplification during merger will likely source the bulk of the magnetic field strength \cite{Kiuchi:2015sga,Palenzuela:2021gdo,Chabanov:2022twz}. After merger, this field produced largely at the interface of the two merging stars will redistribute itself following large scale flow patterns in the remnant \cite{Kastaun:2016yaf}. During all these processes, helicity will be conserved except on smallest scales where numerical resistivity can act. This implies that the merger remnant should contain large scale helicity currents. In the following, we aim to provide a first demonstration of this.

Our simulations have insufficient numerical resolution to fully capture the magnetic field amplification dynamics described above. At the same time, large scale stretching, winding and braking dynamics \cite{Shapiro:2000zh} are fully captured, as are flow structures since these happen on macroscopic scales. This means that while our value of helicity will be off (since it is produced at the resistive scale), the bulk transport flows should be meaningful.

\begin{figure*}
    \centering
    \includegraphics[width=0.95\linewidth]{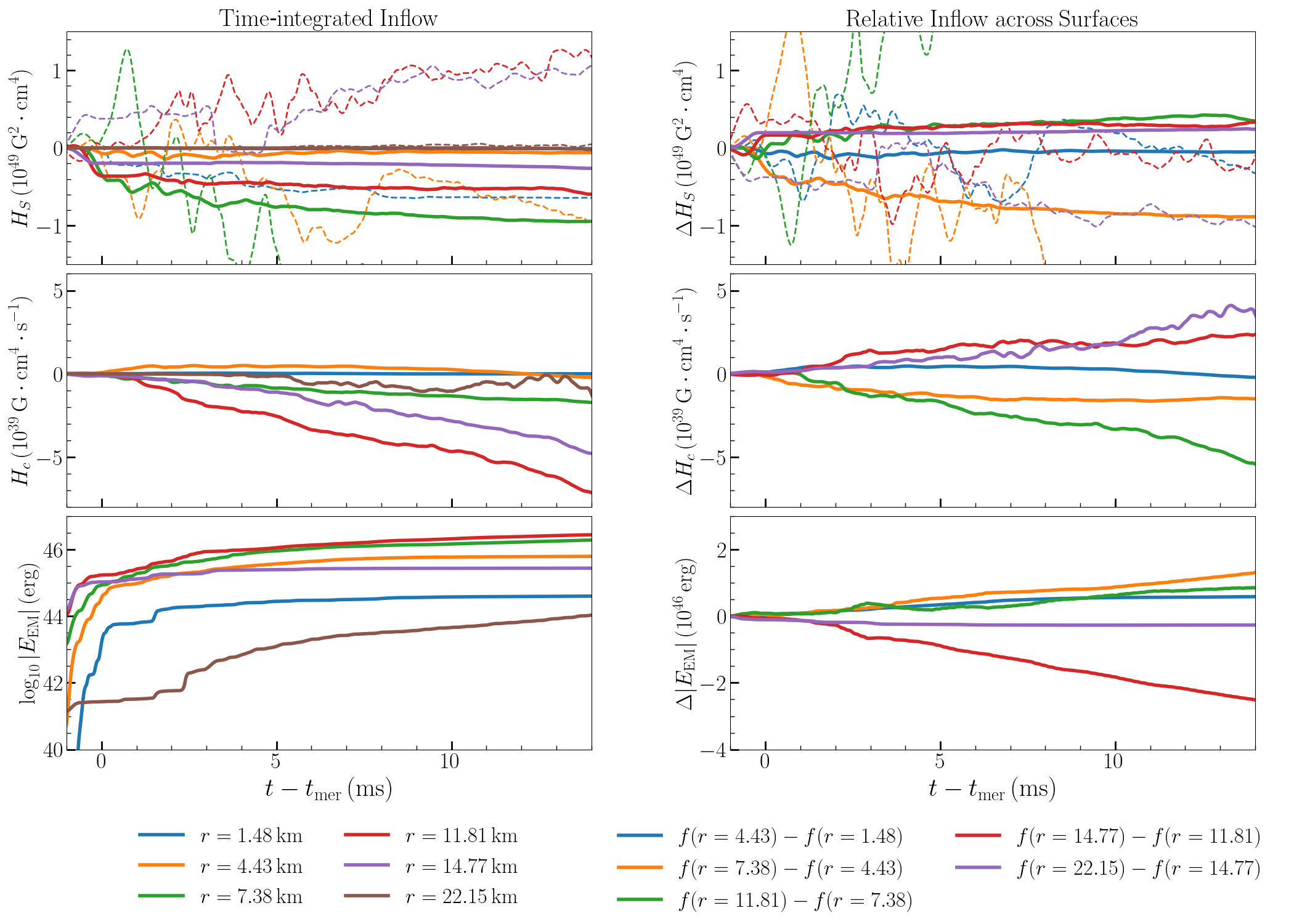}
    \caption{Left column: Cumulative (time-integrated) inflow of magnetic helicity, $H_S$, cross helicity, $H_c$, and electromagnetic energy, $E_{\rm EM}$, over spherical surfaces of radius $r/M_\odot=1, 3, 5, 8, 10, 15$, respectively. Right column: relative differences of the curves on the left by subtracting between inflows of adjacent surfaces. Dashed lines show the gauge-dependent surface magnetic helicity, $H_{S,\Phi}$.}
    \label{fig:hel_diff}
\end{figure*}

In order to track how helicity evolves throughout the merger, we introduce a set of nested spheres centered on the origin of the domain. Since we consider and equal mass merger (see Appendix \ref{app:numerics} for details) this is where the center of mass of the remnant will be. We then compute helicity fluxes through these concentric shells (see left column of Fig. \ref{fig:hel_diff}).

We present the results of these simulations in Fig. \ref{fig:2D_plot}.
We can see that the magnetic field after merger shows turbulent structures in both the star and the disk. At the same time we also show the gauge-dependent magnetic helicity density $\mathcal{H}$ normalized to the background field strength $B$. We can see that in the mixed poloidal/toroidal field geometry, we start with a net helicity density inside the two stars. This helicity is then transported through merger. Finally we can see that both the stellar remnant and the disk have substantial turbulent helicity patches. However, due to the gauge dependence of this quantity, it is difficult to interpret the precise meaning of these patches. Using our gauge-invariant transport formulation, we will provide a more detailed discussion later in this Section, see also Fig. \ref{fig:hel_diff}. As an additional diagnostic quantity, we also track the magnetic cross helicity \cite{Bekenstein87}.
Cross-helicity is defined via
\begin{align}
    \mathcal{H}_c^\mu &= h_T b^\mu\,,\\
    \nabla_\mu \mathcal{H}_c^\mu &= T b^\mu \nabla_\mu s\,, \label{eqn:cross}
\end{align}
where $b^\mu = \left(B^\mu + B^\nu u_\nu u^\mu\right)/(- n_\kappa u^\kappa)$ is the comoving magnetic field, $h_T$ is the specific enthalpy, $s$ being the entropy ber baryon and $T$ being the temperature. Cross-helicity measures the amount of linkage of fluid streamlines with magnetic field lines \cite{Blackman:2014kxa}.

We can see that the merger produces substantial cross-helicity throughout the remnant and disk with similar turbulent patches as we found for the magnetic helicity density (see also Fig. \ref{fig:2D_plot_meridional}).
Since the cross-helicity density does not suffer from gauge ambiguities, we will loosely use it as a reference quantity to contrast it with the magnetic helicity evolution.

From Eq. \eqref{eqn:cross} we can see that cross-helicity is conserved for isentropic flows, and along fluid stream lines with constant entropy.
In the merger, entropy is mainly produced during the collision with the flow patterns in the remnant being largely isentropic \cite{Kastaun:2016yaf,Most:2022wgo}. It is therefore meaningful to additionally consider transport of volume cross-helicity, $H_c$, through the remnant
\begin{align}
    \partial_t H_c =\partial_t \int_V {\rm d}x^3  \sqrt{\gamma} \alpha \mathcal{H}_c^0   \approx - \oint_{\partial V} {\rm d}S\, s_i \sqrt{\gamma} \alpha h_T b^i\,.
\end{align}
This expression parallels that of surface magnetic helicity transport \eqref{eq:surface_transport}, but without the need for an elliptic equation. 

We are now in a position to demonstrate the helicity transport framework presented in this paper. To this end, we calculate the magnetic helicity and cross helicity fluxes on surfaces of constant radius $r/M_{\odot} = [1,3,5,8,10,15]$ from the origin (see Fig. \ref{fig:hel_diff}).  We can see that during merger the helicity begins to move throughout the remnant. We can best see this when comparing the fluxes through consecutive shells (right column, Fig. \ref{fig:hel_diff}). We first see that there is a net outflow of helicity from the center. This is consistent with regions of high temperature (and entropy) rearranging into a torus shape after being initial produced at the collision interface \cite{Kastaun:2016yaf}. Consequently, both cross helicity and magnetic helicity feature a strong inflow into these regions (green curves) and outflow from the inner regions (orange curves).  
The inflow/outflow of helicity also correlates with an outflow/inflow of electromagnetic energy, $E_{\rm EM}$ into these regions. 
Most importantly, we see a clear outflow of helicity into the outer regions of the remnant neutron star, which may affect the presence of large scale $\alpha\Omega-$dynamos in that region \cite{Kiuchi:2023obe,Most:2023sme}.

Overall, this leads to regions of oppositely signed helicity inside the merger remnant. To assess the importance of a gauge-independent helicity calculation, we now compare the flux of magnetic helicity using definitions for both gauge-independent \eqref{eq:4d_rel_hel} and gauge-dependent \eqref{eq:4d_hel_def} transport. By following the procedure outlined in section \ref{sec:GREM}, the gauge-dependent helicity flux is given by
\begin{equation}
    H_{S,\Phi}=\oint_{\partial V}{\text{d}Ss_i\sqrt{\gamma}(\alpha\Phi B^i-\varepsilon^{ijk}A_jD_k)}.
\end{equation}
Note that now the scalar potential, $\Phi$, contribution does not vanish. The result is shown with dashed lines in Fig.~\ref{fig:hel_diff}. We can see that the gauge-dependent flux is highly oscillatory, changes sign frequently, and is even inconsistent sometimes (e.g., red curve). Clearly, a gauge-independent treatment is important if any conclusion about helicity transport is to be drawn.

\section{Discussion}\label{sec:conclusions}

Magnetic helicity is an essential quantity of magnetohydrodynamic flows and dynamos. For nonresistive plasmas present in many relativistic astrophysical systems it is conserved. As a topological invariant, it can affect the formation of large scale magnetic fields \cite{Blackman:2014kxa}, as well as the feasibility of some mean-field dynamo models \cite{Gruzinov:1994zz}.

Magnetic helicity follows a continuity-type transport equation, but its interpretation is complicated by an apparent lack of gauge invariance. In essence, helicity is only meaningfully defined as a global integral over linked field lines solely confined in a given volume. The need to measure helicity in arbitrary patches of a given system therefore requires separating field lines contained in that volume, from field lines leaving it.

By generalizing the concepts of relative helicity by Berger \& Field \cite{Berger_Field_1984} and Finn \& Antonsen \cite{Finn_Antonsen_1985} to the relativistic context, we have provided such a formulation that is gauge-invariant under electromagnetic and general-relativistic gauge transformations.
In doing so, we have adopted the approach of Ref. \cite{Schuck_2019} proposed for solar plasmas to formulate the transport fluxes as a two-dimensional elliptic problem. These depend only on the fluid velocity and local magnetic field, making the suitable for a broad variety of general-relativistic magnetohydrodynamics (GRMHD) codes that do not evolve the magnetic vector potential \cite{Font:2008fka}.

We have then applied this formulation to a neutron star merger, showing that after merger radial zones of different helicity form, which are associated with global currents redistributing small scale turbulent fields throughout the remnant.

It has recently been suggested that the outer layers of the hypermassive neutron star remnant are subject to strong magnetic field amplification from large-scale $\alpha\Omega-$dynamos \cite{Kiuchi:2023obe}. These could affect break out of the magnetic field from the star and subsequent jet and wind launching \cite{Most:2023sft,Combi:2023yav,Kiuchi:2023obe,Most:2023sme}.
Since the $\alpha\Omega-$dynamo is subject to quenching depending on the local helicity value, it is imperative to understand the helicity background on which it is operating on.
Our results indicate that global helicity currents are present, leading to an inflow of net helicity into these regions. 

We caution that our simulation is insufficient to capture the turbulent field amplification process during merger \cite{Kiuchi:2015sga,Aguilera-Miret:2021fre}. In this sense, the redistribution of helicity we observe is largely from the initial helicity imparted onto the inspiralling neutron stars. In line with large-scale background hydrodynamical flow structures which are well captured at the current numerical resolution, however, our simulation is well equipped to demonstrate the redistribution of magnetic helicity into a ring-shaped structure, in analogy to the rearrangement of hot regions in the early post-merger phase \cite{Kastaun:2016yaf}. As such, the large-scale magnetic helicity currents we identify may critically affect the existence of helical dynamo processes.

While our application to neutron star mergers has so far been mainly a proof-of-concept, we expect our formulation of helicity transport to be especially relevant for high resolution global simulations of dynamo action in neutron star mergers and black hole accretion disks. 
We plan to carry out such studies and analysis in future work.

\begin{acknowledgments}
The authors are grateful to N. Vu for helpful advice concerning numerical solutions of elliptic problems.
ERM is grateful for insightful discussions with J. Beattie, A. Bhattacharjee, C. Palenzuela, A. Philippov, D. Radice, E. Vishniac and Y. Zenati.
ERM acknowledges support by the National Science Foundation under grant No. PHY-2309210.  This work mainly used Delta at the National Center for Supercomputing Applications (NCSA) through allocation PHY210074 from the Advanced Cyberinfrastructure Coordination Ecosystem: Services \& Support (ACCESS) program, which is supported by National Science Foundation grants \#2138259, \#2138286, \#2138307, \#2137603, and \#2138296.  Additional simulations were performed on the NSF Frontera supercomputer under grant AST21006.
ERM gratefully acknowledges discussions and participation at a workshop at the Kavli Institute for Theoretical Physics. This research was supported in part by grant NSF PHY-2309135 to the Kavli Institute for Theoretical Physics (KITP).
\end{acknowledgments}

\appendix

\section{Decomposition of the reference field}
\label{section:Append_decomp}
The Helmholtz decomposition theorem (see Ref. \cite{Schuck_2019} for an extensive discussion) guarantees that any finite, integrable, and continuously differentiable vector function defined in a simply connected volume can be uniquely decomposed into a solenoidal component (i.e. a curl of vector function) plus an irrotational component (i.e. a gradient of scalar). Performing such a decomposition with respect to the vector field $\boldsymbol{D}_R$ results in Eq.~\eqref{eq:first_decomp} with boundary conditions Eqs.~\eqref{eq:helmholtz_decomp_bd1} and \eqref{eq:helmholtz_decomp_bd2}. Thus, Eq.~\eqref{eq:hel_evol_DB} turns into
\begin{eqnarray}
    &&\int_V{\text{d}^3x\,\frac{\partial}{\partial t}(\alpha\sqrt{\gamma}h^0)}\notag\\
    &=&-2\int_V{\text{d}^3x\,\sqrt{\gamma}D_iB^i}+2\int_V{\text{d}^3x\,\sqrt{\gamma}(\Sigma_{Ri}+\mathcal{D}_i\Lambda_R)B_R^i}\notag\\
    &-&\oint_{\partial V}{\text{d}S\, s_i\sqrt{\gamma}\varepsilon^{ijk}(A_j+A_{Rj})(D_k-\Sigma_{Rk}-\mathcal{D}_k\Lambda_R) }\label{eq:int_time_evol_decom}
\end{eqnarray}

The second term on the RHS of Eq.~\eqref{eq:int_time_evol_decom} is
\begin{eqnarray}
    &&\int_V{\text{d}^3x\,\sqrt{\gamma}(\Sigma_{Ri}+\mathcal{D}_i\Lambda_R)B_R^i}\notag\\
    &=&-\int_V{\text{d}^3x\,\sqrt{\gamma}\Sigma_{Ri}\mathcal{D}^i\psi_R}+\int_V{\text{d}^3x\,\sqrt{\gamma}\mathcal{D}_i\Lambda_RB_R^i}\notag\\
    &=&\int_V{\text{d}^3x\,\sqrt{\gamma}\left[\psi_R\mathcal{D}^i\Sigma_{Ri}-\mathcal{D}^i(\psi_R\Sigma_{Ri})\right]}\notag\\
    &&+\int_V{\text{d}^3x\,\sqrt{\gamma}\left[\mathcal{D}_i(\Lambda_RB_R^i)-\Lambda_R\mathcal{D}_iB_R^i\right]}\notag\\
    &=&\oint_{\partial V}{\text{d}S\sqrt{\gamma} s_i(\psi_R\Sigma_R^i-\Lambda_RB_R^i)}\notag\\
    &=&\Lambda_R|_S\int_V{\text{d}^3x\,\sqrt{\gamma}\mathcal{D}_iB_R^i}=0.
\end{eqnarray}
where we use the boundary conditions in Eqs.~\eqref{eq:ref_fd_bd1},\,\eqref{eq:helmholtz_decomp_bd1}, and \eqref{eq:helmholtz_decomp_bd2}. We can also show that the third term on the RHS of Eq.~\eqref{eq:int_time_evol_decom} can be simplified as $\oint_{\partial V}{\text{d}S\, s_i\sqrt{\gamma}\varepsilon^{ijk}(A_j+A_{Rj})(D_k-\Sigma_{Rk})}$, since $\mathcal{D}_k\Lambda_R\propto s_k$ due to its constancy on the boundary and the antisymmetric feature of $\varepsilon^{ijk} s_i s_k$. After this decomposition, we reach Eq.~\eqref{eq:hel_evol_1st_simp}

\section{Gauge invariance of relative helicity transport}\label{sec:gauge_invariance}

Next we show the gauge invariance of Eq.~\eqref{eq:hel_evol_1st_simp}. We add a guage transformation to the vector potential
\begin{equation}
    A_j\rightarrow A_j+\partial_j\Lambda \label{eq:gauge_transform}
\end{equation}

The first term on the RHS of Eq.~\eqref{eq:hel_evol_1st_simp}, which represent the magnetic helicity inside the probing volume, are automatically gauge-invariant. The second term, which is the surface transport ($\frac{\partial \mathcal{H}_S}{\partial t}$), can also be shown to be gauge-invariant. For simplicity, we still use $D_k-D_{Rk}$ rather than $D_k-\Sigma_{Rk}$ to verify the gauge invariance.
\begin{eqnarray}
    \frac{\partial H_S'}{\partial t}&=&\frac{\partial H_S}{\partial t}-\oint_{\partial V}{\text{d}S\,s_i\sqrt{\gamma}\varepsilon^{ijk}\partial_j\Lambda(D_k-D_{Rk})}\notag\\
    &=&\frac{\partial H_S}{\partial t}-\oint_{\partial V}{\text{d}S\,s_i\sqrt{\gamma}\varepsilon^{ijk}\partial_j[\Lambda(D_k-D_{Rk})]}\notag\\
    &&+\oint_{\partial V}{\text{d}S\,s_i\sqrt{\gamma}\varepsilon^{ijk}\Lambda\partial_j(D_k-D_{Rk})}\notag\\
    &=&\frac{\partial H_S}{\partial t}+\int_V{\text{d}^3x\,\epsilon^{ijk}\partial_i\partial_j[\Lambda(D_k-D_{Rk})]}\notag\\
    &&-\oint_{\partial V}{\text{d}S\,s_i\Lambda\partial_t\left[\sqrt{\gamma}(B^i-B_R^i)\right]}\notag\\
    &=&\frac{\partial H_S}{\partial t}
\end{eqnarray}
where we use Eq.~\eqref{eq:ref_fd_bd1} and antisymmetry of three-dimensional Levi-Civita tensor.

\section{Helmholtz-Hodge surface decomposition}
\label{app:helmholtz}
The Helmholtz-Hodge decomposition \cite{hodge1989theory} asserts that any finite, square integrable vector function on a $\mathcal{C}^2$ surface, which is satisfied in our problem as spherical surface, can be separated into a normal part plus a surface part, which is further splitted into a solenoidal component and a lamallar (i.e. whose surface curl resides in the tangent space of the surface) component. The decomposition is presented in Eqs.~\eqref{eq:helmholtz-hodge_decomp} and \eqref{eq:helmholtz-hodge_decomp_ref} with the three components determined by Eqs.~\eqref{eq:norm_comp}-\eqref{eq:PDE_of_zeta}. Below we show a proof of Eq.~\eqref{eq:hel_evol_2nd_simp}

First, we calculate a relation between the solenoidal parts $\chi$ and $\chi_R$. Using the definition of $\chi$ in Eq.~\eqref{eq:PDE_of_chi}, we find
\begin{eqnarray}
    \partial_S^2\partial_t\chi&=&-s_i\varepsilon^{ijk}\partial_{Sj}(D_k-\tau s_k)\notag\\
    &=&-s_i\varepsilon^{ijk}(\partial_j-s_j s^l\partial_l)(D_k-\tau s_k)\notag\\
    &=&-s_i\varepsilon^{ijk}\partial_j D_k+s_i\varepsilon^{ijk}\partial_j(\tau s_k)\notag\\
    &=&s_i\frac{1}{\sqrt{\gamma}}\partial_t(\sqrt{\gamma}B^i)+\varepsilon^{ijk}s_i s_k\partial_j\tau+\tau s_i\varepsilon^{ijk}\partial_j s_k\notag\\
    &=&\frac{1}{\sqrt{\gamma}}\partial_t(\sqrt{\gamma}s_iB^i),\label{eq:chi_simp}
\end{eqnarray}
where we use the fact that the normal vector, $s^i$, is irrotational.

Similarly, for $\chi_R$ we have
\begin{equation}
    \partial_S^2\partial_t\chi_R=\frac{1}{\sqrt{\gamma}}\partial_t(\sqrt{\gamma}s_iB_R^i).
\end{equation}
Combining the two equation together and using the boundary condition Eq.~\eqref{eq:ref_fd_bd1}, we conclude that
\begin{equation}
    \partial_S^2\left(\frac{\partial\chi_P}{\partial t}-\frac{\partial\chi}{\partial t}\right)=0
\end{equation}
which means
\begin{equation}
    \frac{\partial\chi}{\partial t}\equiv\frac{\partial\chi_P}{\partial t}+C \label{eq:chirelation}
\end{equation}
on boundary $S$, where $C$ is a constant.

We now substitute the decomposition of Eq.~\eqref{eq:helmholtz-hodge_decomp} and \eqref{eq:helmholtz-hodge_decomp_ref} into the surface helicity transport term in Eq.~\eqref{eq:hel_evol_2nd_simp} and find
\begin{eqnarray}
    &&\frac{\partial H_S}{\partial t}=-\oint_{\partial V}{\text{d}S\, s_i\sqrt{\gamma}\varepsilon^{ijk}(A_j+A_{Rj})(D_k-\Sigma_{Rk})}\notag\\
    &=&-\oint_{\partial V}{\text{d}S\,s_i\sqrt{\gamma}\varepsilon^{ijk}(A_j+A_{Rj})(\tau-\tau_R) s_k}\notag\\
    &&+\oint_{\partial V}{\text{d}S\,s_i\sqrt{\gamma}\varepsilon^{ijk}(A_j+A_{Rj})}\times\notag\\
    &&\qquad\qquad\qquad\varepsilon_{klm} s^l(\partial^m-s^m s_n\partial^n)\left(\frac{\partial\chi}{\partial t}-\frac{\partial\chi_R}{\partial t}\right)\notag\\
    &&+\oint_{\partial V}{\text{d}S\,s_i\sqrt{\gamma}\varepsilon^{ijk}(A_j+A_{Rj})(\partial_k-s_k s_l\partial^l)(\zeta-\zeta_R)}\notag\\
    &=&\oint_{\partial V}{\text{d}S\,s_i\sqrt{\gamma}\varepsilon^{ijk}(A_j+A_{Rj})\partial_k(\zeta-\zeta_R)} \label{eq:surface_simp}
\end{eqnarray}
The first term on the right-hand-side of the second equation vanishes due to antisymmetric $\varepsilon^{ijk} s_i s_k$, and the second term also vanishes bacause Eq.~\eqref{eq:chirelation} ensures that the surface divergence operator $\partial^m-s^m s_n\partial^n$ produces zero when acting on a surface constant object. Further, we can show
\begin{eqnarray}
    \frac{\partial H_S}{\partial t}&=&-\oint_{\partial V}{\text{d}S\,s_i \sqrt{\gamma}\varepsilon^{ijk}(\zeta-\zeta_R)\partial_k (A_j+A_{Rj})}\notag\\
    &&+\oint_{\partial V}{\text{d}S\,s_i\sqrt{\gamma} \varepsilon^{ijk}\partial_k [(A_j+A_{Rj})(\zeta-\zeta_R)]}\notag\\
    &=&\oint_{\partial V}{\text{d}S\,\sqrt{\gamma}(\zeta-\zeta_R) s_i (B^i+B_R^i)}\notag\\
    &&-\int_V{\text{d}^3x\,\sqrt{\gamma}\varepsilon^{ijk}\partial_i\partial_k[(A_j+A_{Rj})(\zeta-\zeta_R)]}\notag\\
    &=&2\oint_{\partial V}{\text{d}S\,\sqrt{\gamma} s_i B^i(\zeta-\zeta_R)} \label{eq:sf_hlc_trsp}
\end{eqnarray}
where we use the boundary condition Eq.~\eqref{eq:ref_fd_bd1} in the last equation.

\section{Numerical relativity simulations}\label{app:numerics}

We numerically solve the coupled Einstein-GRMHD system to evolve the dynamical merger phase of a binary neutron star system.
We do so by evolving the spacetime dynamics using the Z4c formulation of the Einstein equations \cite{Bernuzzi:2009ex,Hilditch:2012fp} in moving puncture gauge \cite{Alcubierre:2002kk}. Furthermore, we solve the GRMHD equations in dynamical spacetimes \cite{Duez:2005sf} using a vector potential formulation in Lorenz gauge \cite{Etienne:2010ui,Etienne:2011re}. This allows us to have direct access to the vector potential and to compute local quantities like the gauge-dependent helicity density, $\mathcal{H}$. \\
We solve the above system using the \texttt{Frankfurt/IllinoisGRMHD (FIL)} code \cite{Most:2019kfe,Etienne:2015cea}. \texttt{FIL} utilizes the \texttt{EinsteinToolkit} infrastructure \cite{Loffler:2011ay}, and implements fourth order unlimited finite-difference for the spacetime \cite{Zlochower:2005bj}, as well as conservative finite difference for the GRMHD sector following the ECHO scheme \cite{DelZanna:2007pk}.
The initial data for the binary system is chosen to be an equal mass binary with a total mass of $M = 2.7\, M_\odot$ using the DD2 equation of state \cite{Hempel:2009mc}. The initial data is computed using the spectral \texttt{Kadath}\cite{Grandclement:2009ju}/\texttt{FUKA}\cite{Papenfort:2021hod} framework.
Our simulations use a nested domain with 8 refinement levels (see Ref. \cite{Most:2021ktk} for details), and a finest resolution of $\Delta x= 260 \, \rm m$. 
The magnetic field geometry is initialized as a mixed poloidal-toroidal field, with $A_\phi= A^0_\phi \max \left(p-0.04\, p_{\rm max}\right)^2$, $A_z= A^0_z \max \left(p-0.04\, p_{\rm max}\right)^2$, where $p$ is the fluid pressure and $p_{\rm max}$ its maximum value inside each star. We choose the overall normalization such that we start out with a maximum of $B_{\rm max} = 10^{15}\, \rm G$ inside the star, as well as a ratio of $A^0_z/A^0_\phi = 0.05$ of initial toroidal to poloidal field.

\newcommand\aj{\ref@jnl{AJ}}%        % Astronomical Journal
\newcommand\psj{\ref@jnl{PSJ}}%       % Planetary Science Journal
\newcommand\araa{\ref@jnl{ARA\&A}}%  % Annual Review of Astron and Astrophys
\renewcommand\apj{\ref@jnl{ApJ}}%    % Astrophysical Journal
\newcommand\apjl{\ref@jnl{ApJL}}     % Astrophysical Journal, Letters
\newcommand\apjs{\ref@jnl{ApJS}}%    % Astrophysical Journal, Supplement
\renewcommand\ao{\ref@jnl{ApOpt}}%   % Applied Optics
\newcommand\apss{\ref@jnl{Ap\&SS}}%  % Astrophysics and Space Science
\newcommand\aap{\ref@jnl{A\&A}}%     % Astronomy and Astrophysics
\newcommand\aapr{\ref@jnl{A\&A~Rv}}%  % Astronomy and Astrophysics Reviews
\newcommand\aaps{\ref@jnl{A\&AS}}%    % Astronomy and Astrophysics, Supplement
\newcommand\azh{\ref@jnl{AZh}}%       % Astronomicheskii Zhurnal
\newcommand\baas{\ref@jnl{BAAS}}%     % Bulletin of the AAS
\newcommand\icarus{\ref@jnl{Icarus}}% % Icarus
\newcommand\jaavso{\ref@jnl{JAAVSO}}  % The Journal of the American Association of Variable Star Observers
\newcommand\jrasc{\ref@jnl{JRASC}}%   % Journal of the RAS of Canada
\newcommand\memras{\ref@jnl{MmRAS}}%  % Memoirs of the RAS
\newcommand\mnras{\ref@jnl{MNRAS}}%   % Monthly Notices of the RAS
\renewcommand\pra{\ref@jnl{PhRvA}}% % Physical Review A: General Physics
\renewcommand\prb{\ref@jnl{PhRvB}}% % Physical Review B: Solid State
\renewcommand\prc{\ref@jnl{PhRvC}}% % Physical Review C
\renewcommand\prd{\ref@jnl{PhRvD}}% % Physical Review D
\renewcommand\pre{\ref@jnl{PhRvE}}% % Physical Review E
\renewcommand\prl{\ref@jnl{PhRvL}}% % Physical Review Letters
\newcommand\pasp{\ref@jnl{PASP}}%     % Publications of the ASP
\newcommand\pasj{\ref@jnl{PASJ}}%     % Publications of the ASJ
\newcommand\qjras{\ref@jnl{QJRAS}}%   % Quarterly Journal of the RAS
\newcommand\skytel{\ref@jnl{S\&T}}%   % Sky and Telescope
\newcommand\solphys{\ref@jnl{SoPh}}% % Solar Physics
\newcommand\sovast{\ref@jnl{Soviet~Ast.}}% % Soviet Astronomy
\newcommand\ssr{\ref@jnl{SSRv}}% % Space Science Reviews
\newcommand\zap{\ref@jnl{ZA}}%       % Zeitschrift fuer Astrophysik
\renewcommand\nat{\ref@jnl{Nature}}%  % Nature
\newcommand\iaucirc{\ref@jnl{IAUC}}% % IAU Cirulars
\newcommand\aplett{\ref@jnl{Astrophys.~Lett.}}%  % Astrophysics Letters
\newcommand\apspr{\ref@jnl{Astrophys.~Space~Phys.~Res.}}% % Astrophysics Space Physics Research
\newcommand\bain{\ref@jnl{BAN}}% % Bulletin Astronomical Institute of the Netherlands
\newcommand\fcp{\ref@jnl{FCPh}}%   % Fundamental Cosmic Physics
\newcommand\gca{\ref@jnl{GeoCoA}}% % Geochimica Cosmochimica Acta
\newcommand\grl{\ref@jnl{Geophys.~Res.~Lett.}}%  % Geophysics Research Letters
\renewcommand\jcp{\ref@jnl{JChPh}}%     % Journal of Chemical Physics
\newcommand\jgr{\ref@jnl{J.~Geophys.~Res.}}%     % Journal of Geophysics Research
\newcommand\jqsrt{\ref@jnl{JQSRT}}%   % Journal of Quantitiative Spectroscopy and Radiative Trasfer
\newcommand\memsai{\ref@jnl{MmSAI}}% % Mem. Societa Astronomica Italiana
\newcommand\nphysa{\ref@jnl{NuPhA}}%     % Nuclear Physics A
\newcommand\physrep{\ref@jnl{PhR}}%       % Physics Reports
\newcommand\physscr{\ref@jnl{PhyS}}%        % Physica Scripta
\newcommand\planss{\ref@jnl{Planet.~Space~Sci.}}%  % Planetary Space Science
\newcommand\procspie{\ref@jnl{Proc.~SPIE}}%      % Proceedings of the SPIE

\newcommand\actaa{\ref@jnl{AcA}}%  % Acta Astronomica
\newcommand\caa{\ref@jnl{ChA\&A}}%  % Chinese Astronomy and Astrophysics
\newcommand\cjaa{\ref@jnl{ChJA\&A}}%  % Chinese Journal of Astronomy and Astrophysics
\newcommand\jcap{\ref@jnl{JCAP}}%  % Journal of Cosmology and Astroparticle Physics
\newcommand\na{\ref@jnl{NewA}}%  % New Astronomy
\newcommand\nar{\ref@jnl{NewAR}}%  % New Astronomy Review
\newcommand\pasa{\ref@jnl{PASA}}%  % Publications of the Astron. Soc. of Australia
\newcommand\rmxaa{\ref@jnl{RMxAA}}%  % Revista Mexicana de Astronomia y Astrofisica

\bibliography{reference}

\begin{thebibliography}{64}%
\makeatletter
\providecommand \@ifxundefined [1]{%
 \@ifx{#1\undefined}
}%
\providecommand \@ifnum [1]{%
 \ifnum #1\expandafter \@firstoftwo
 \else \expandafter \@secondoftwo
 \fi
}%
\providecommand \@ifx [1]{%
 \ifx #1\expandafter \@firstoftwo
 \else \expandafter \@secondoftwo
 \fi
}%
\providecommand \natexlab [1]{#1}%
\providecommand \enquote  [1]{``#1''}%
\providecommand \bibnamefont  [1]{#1}%
\providecommand \bibfnamefont [1]{#1}%
\providecommand \citenamefont [1]{#1}%
\providecommand \href@noop [0]{\@secondoftwo}%
\providecommand \href [0]{\begingroup \@sanitize@url \@href}%
\providecommand \@href[1]{\@@startlink{#1}\@@href}%
\providecommand \@@href[1]{\endgroup#1\@@endlink}%
\providecommand \@sanitize@url [0]{\catcode `\\12\catcode `\$12\catcode `\&12\catcode `\#12\catcode `\^12\catcode `\_12\catcode `\%12\relax}%
\providecommand \@@startlink[1]{}%
\providecommand \@@endlink[0]{}%
\providecommand \url  [0]{\begingroup\@sanitize@url \@url }%
\providecommand \@url [1]{\endgroup\@href {#1}{\urlprefix }}%
\providecommand \urlprefix  [0]{URL }%
\providecommand \Eprint [0]{\href }%
\providecommand \doibase [0]{https://doi.org/}%
\providecommand \selectlanguage [0]{\@gobble}%
\providecommand \bibinfo  [0]{\@secondoftwo}%
\providecommand \bibfield  [0]{\@secondoftwo}%
\providecommand \translation [1]{[#1]}%
\providecommand \BibitemOpen [0]{}%
\providecommand \bibitemStop [0]{}%
\providecommand \bibitemNoStop [0]{.\EOS\space}%
\providecommand \EOS [0]{\spacefactor3000\relax}%
\providecommand \BibitemShut  [1]{\csname bibitem#1\endcsname}%
\let\auto@bib@innerbib\@empty
%</preamble>
\bibitem [{\citenamefont {Charbonneau}(2014)}]{annurev:/content/journals/10.1146/annurev-astro-081913-040012}%
  \BibitemOpen
  \bibfield  {author} {\bibinfo {author} {\bibfnamefont {P.}~\bibnamefont {Charbonneau}},\ }\bibfield  {title} {\bibinfo {title} {Solar dynamo theory},\ }\href {https://doi.org/https://doi.org/10.1146/annurev-astro-081913-040012} {\bibfield  {journal} {\bibinfo  {journal} {Annual Review of Astronomy and Astrophysics}\ }\textbf {\bibinfo {volume} {52}},\ \bibinfo {pages} {251} (\bibinfo {year} {2014})}\BibitemShut {NoStop}%
\bibitem [{\citenamefont {{Ruediger}}\ \emph {et~al.}(1995)\citenamefont {{Ruediger}}, \citenamefont {{Elstner}},\ and\ \citenamefont {{Stepinski}}}]{1995A&A...298..934R}%
  \BibitemOpen
  \bibfield  {author} {\bibinfo {author} {\bibfnamefont {G.}~\bibnamefont {{Ruediger}}}, \bibinfo {author} {\bibfnamefont {D.}~\bibnamefont {{Elstner}}},\ and\ \bibinfo {author} {\bibfnamefont {T.~F.}\ \bibnamefont {{Stepinski}}},\ }\bibfield  {title} {\bibinfo {title} {{The standard-accretion disk dynamo.}},\ }\href@noop {} {\bibfield  {journal} {\bibinfo  {journal} {Astron. \& Astrophys.}\ }\textbf {\bibinfo {volume} {298}},\ \bibinfo {pages} {934} (\bibinfo {year} {1995})}\BibitemShut {NoStop}%
\bibitem [{\citenamefont {Roberts}\ and\ \citenamefont {Glatzmaier}(2000)}]{RevModPhys.72.1081}%
  \BibitemOpen
  \bibfield  {author} {\bibinfo {author} {\bibfnamefont {P.~H.}\ \bibnamefont {Roberts}}\ and\ \bibinfo {author} {\bibfnamefont {G.~A.}\ \bibnamefont {Glatzmaier}},\ }\bibfield  {title} {\bibinfo {title} {Geodynamo theory and simulations},\ }\href {https://doi.org/10.1103/RevModPhys.72.1081} {\bibfield  {journal} {\bibinfo  {journal} {Rev. Mod. Phys.}\ }\textbf {\bibinfo {volume} {72}},\ \bibinfo {pages} {1081} (\bibinfo {year} {2000})}\BibitemShut {NoStop}%
\bibitem [{\citenamefont {Brandenburg}\ and\ \citenamefont {Ntormousi}(2023)}]{annurev:/content/journals/10.1146/annurev-astro-071221-052807}%
  \BibitemOpen
  \bibfield  {author} {\bibinfo {author} {\bibfnamefont {A.}~\bibnamefont {Brandenburg}}\ and\ \bibinfo {author} {\bibfnamefont {E.}~\bibnamefont {Ntormousi}},\ }\bibfield  {title} {\bibinfo {title} {Galactic dynamos},\ }\href {https://doi.org/https://doi.org/10.1146/annurev-astro-071221-052807} {\bibfield  {journal} {\bibinfo  {journal} {Annual Review of Astronomy and Astrophysics}\ }\textbf {\bibinfo {volume} {61}},\ \bibinfo {pages} {561} (\bibinfo {year} {2023})}\BibitemShut {NoStop}%
\bibitem [{\citenamefont {Brandenburg}\ \emph {et~al.}(2012)\citenamefont {Brandenburg}, \citenamefont {Sokoloff},\ and\ \citenamefont {Subramanian}}]{brandenburg2012current}%
  \BibitemOpen
  \bibfield  {author} {\bibinfo {author} {\bibfnamefont {A.}~\bibnamefont {Brandenburg}}, \bibinfo {author} {\bibfnamefont {D.}~\bibnamefont {Sokoloff}},\ and\ \bibinfo {author} {\bibfnamefont {K.}~\bibnamefont {Subramanian}},\ }\bibfield  {title} {\bibinfo {title} {Current status of turbulent dynamo theory: from large-scale to small-scale dynamos},\ }\href@noop {} {\bibfield  {journal} {\bibinfo  {journal} {Space Science Reviews}\ }\textbf {\bibinfo {volume} {169}},\ \bibinfo {pages} {123} (\bibinfo {year} {2012})}\BibitemShut {NoStop}%
\bibitem [{\citenamefont {Schekochihin}(2022)}]{Schekochihin_2022}%
  \BibitemOpen
  \bibfield  {author} {\bibinfo {author} {\bibfnamefont {A.~A.}\ \bibnamefont {Schekochihin}},\ }\bibfield  {title} {\bibinfo {title} {Mhd turbulence: a biased review},\ }\href {https://doi.org/10.1017/S0022377822000721} {\bibfield  {journal} {\bibinfo  {journal} {Journal of Plasma Physics}\ }\textbf {\bibinfo {volume} {88}},\ \bibinfo {pages} {155880501} (\bibinfo {year} {2022})}\BibitemShut {NoStop}%
\bibitem [{\citenamefont {Kiuchi}\ \emph {et~al.}(2018)\citenamefont {Kiuchi}, \citenamefont {Kyutoku}, \citenamefont {Sekiguchi},\ and\ \citenamefont {Shibata}}]{Kiuchi:2017zzg}%
  \BibitemOpen
  \bibfield  {author} {\bibinfo {author} {\bibfnamefont {K.}~\bibnamefont {Kiuchi}}, \bibinfo {author} {\bibfnamefont {K.}~\bibnamefont {Kyutoku}}, \bibinfo {author} {\bibfnamefont {Y.}~\bibnamefont {Sekiguchi}},\ and\ \bibinfo {author} {\bibfnamefont {M.}~\bibnamefont {Shibata}},\ }\bibfield  {title} {\bibinfo {title} {{Global simulations of strongly magnetized remnant massive neutron stars formed in binary neutron star mergers}},\ }\href {https://doi.org/10.1103/PhysRevD.97.124039} {\bibfield  {journal} {\bibinfo  {journal} {Phys. Rev. D}\ }\textbf {\bibinfo {volume} {97}},\ \bibinfo {pages} {124039} (\bibinfo {year} {2018})},\ \Eprint {https://arxiv.org/abs/1710.01311} {arXiv:1710.01311 [astro-ph.HE]} \BibitemShut {NoStop}%
\bibitem [{\citenamefont {Kiuchi}\ \emph {et~al.}(2024)\citenamefont {Kiuchi}, \citenamefont {Reboul-Salze}, \citenamefont {Shibata},\ and\ \citenamefont {Sekiguchi}}]{Kiuchi:2023obe}%
  \BibitemOpen
  \bibfield  {author} {\bibinfo {author} {\bibfnamefont {K.}~\bibnamefont {Kiuchi}}, \bibinfo {author} {\bibfnamefont {A.}~\bibnamefont {Reboul-Salze}}, \bibinfo {author} {\bibfnamefont {M.}~\bibnamefont {Shibata}},\ and\ \bibinfo {author} {\bibfnamefont {Y.}~\bibnamefont {Sekiguchi}},\ }\bibfield  {title} {\bibinfo {title} {{A large-scale magnetic field produced by a solar-like dynamo in binary neutron star mergers}},\ }\href {https://doi.org/10.1038/s41550-024-02194-y} {\bibfield  {journal} {\bibinfo  {journal} {Nature Astron.}\ }\textbf {\bibinfo {volume} {8}},\ \bibinfo {pages} {298} (\bibinfo {year} {2024})},\ \Eprint {https://arxiv.org/abs/2306.15721} {arXiv:2306.15721 [astro-ph.HE]} \BibitemShut {NoStop}%
\bibitem [{\citenamefont {Combi}\ and\ \citenamefont {Siegel}(2023)}]{Combi:2023yav}%
  \BibitemOpen
  \bibfield  {author} {\bibinfo {author} {\bibfnamefont {L.}~\bibnamefont {Combi}}\ and\ \bibinfo {author} {\bibfnamefont {D.~M.}\ \bibnamefont {Siegel}},\ }\bibfield  {title} {\bibinfo {title} {{Jets from Neutron-Star Merger Remnants and Massive Blue Kilonovae}},\ }\href {https://doi.org/10.1103/PhysRevLett.131.231402} {\bibfield  {journal} {\bibinfo  {journal} {Phys. Rev. Lett.}\ }\textbf {\bibinfo {volume} {131}},\ \bibinfo {pages} {231402} (\bibinfo {year} {2023})},\ \Eprint {https://arxiv.org/abs/2303.12284} {arXiv:2303.12284 [astro-ph.HE]} \BibitemShut {NoStop}%
\bibitem [{\citenamefont {Most}\ and\ \citenamefont {Quataert}(2023)}]{Most:2023sft}%
  \BibitemOpen
  \bibfield  {author} {\bibinfo {author} {\bibfnamefont {E.~R.}\ \bibnamefont {Most}}\ and\ \bibinfo {author} {\bibfnamefont {E.}~\bibnamefont {Quataert}},\ }\bibfield  {title} {\bibinfo {title} {{Flares, Jets, and Quasiperiodic Outbursts from Neutron Star Merger Remnants}},\ }\href {https://doi.org/10.3847/2041-8213/acca84} {\bibfield  {journal} {\bibinfo  {journal} {Astrophys. J. Lett.}\ }\textbf {\bibinfo {volume} {947}},\ \bibinfo {pages} {L15} (\bibinfo {year} {2023})},\ \Eprint {https://arxiv.org/abs/2303.08062} {arXiv:2303.08062 [astro-ph.HE]} \BibitemShut {NoStop}%
\bibitem [{\citenamefont {Aguilera-Miret}\ \emph {et~al.}(2023)\citenamefont {Aguilera-Miret}, \citenamefont {Palenzuela}, \citenamefont {Carrasco},\ and\ \citenamefont {Vigan\`o}}]{Aguilera-Miret:2023qih}%
  \BibitemOpen
  \bibfield  {author} {\bibinfo {author} {\bibfnamefont {R.}~\bibnamefont {Aguilera-Miret}}, \bibinfo {author} {\bibfnamefont {C.}~\bibnamefont {Palenzuela}}, \bibinfo {author} {\bibfnamefont {F.}~\bibnamefont {Carrasco}},\ and\ \bibinfo {author} {\bibfnamefont {D.}~\bibnamefont {Vigan\`o}},\ }\bibfield  {title} {\bibinfo {title} {{Role of turbulence and winding in the development of large-scale, strong magnetic fields in long-lived remnants of binary neutron star mergers}},\ }\href {https://doi.org/10.1103/PhysRevD.108.103001} {\bibfield  {journal} {\bibinfo  {journal} {Phys. Rev. D}\ }\textbf {\bibinfo {volume} {108}},\ \bibinfo {pages} {103001} (\bibinfo {year} {2023})},\ \Eprint {https://arxiv.org/abs/2307.04837} {arXiv:2307.04837 [astro-ph.HE]} \BibitemShut {NoStop}%
\bibitem [{\citenamefont {Palenzuela}\ \emph {et~al.}(2022)\citenamefont {Palenzuela}, \citenamefont {Aguilera-Miret}, \citenamefont {Carrasco}, \citenamefont {Ciolfi}, \citenamefont {Kalinani}, \citenamefont {Kastaun}, \citenamefont {Mi\~nano},\ and\ \citenamefont {Vigan\`o}}]{Palenzuela:2021gdo}%
  \BibitemOpen
  \bibfield  {author} {\bibinfo {author} {\bibfnamefont {C.}~\bibnamefont {Palenzuela}}, \bibinfo {author} {\bibfnamefont {R.}~\bibnamefont {Aguilera-Miret}}, \bibinfo {author} {\bibfnamefont {F.}~\bibnamefont {Carrasco}}, \bibinfo {author} {\bibfnamefont {R.}~\bibnamefont {Ciolfi}}, \bibinfo {author} {\bibfnamefont {J.~V.}\ \bibnamefont {Kalinani}}, \bibinfo {author} {\bibfnamefont {W.}~\bibnamefont {Kastaun}}, \bibinfo {author} {\bibfnamefont {B.}~\bibnamefont {Mi\~nano}},\ and\ \bibinfo {author} {\bibfnamefont {D.}~\bibnamefont {Vigan\`o}},\ }\bibfield  {title} {\bibinfo {title} {{Turbulent magnetic field amplification in binary neutron star mergers}},\ }\href {https://doi.org/10.1103/PhysRevD.106.023013} {\bibfield  {journal} {\bibinfo  {journal} {Phys. Rev. D}\ }\textbf {\bibinfo {volume} {106}},\ \bibinfo {pages} {023013} (\bibinfo {year} {2022})},\ \Eprint {https://arxiv.org/abs/2112.08413} {arXiv:2112.08413 [gr-qc]} \BibitemShut {NoStop}%
\bibitem [{\citenamefont {Liska}\ \emph {et~al.}(2020)\citenamefont {Liska}, \citenamefont {Tchekhovskoy},\ and\ \citenamefont {Quataert}}]{Liska:2018btr}%
  \BibitemOpen
  \bibfield  {author} {\bibinfo {author} {\bibfnamefont {M.~T.~P.}\ \bibnamefont {Liska}}, \bibinfo {author} {\bibfnamefont {A.}~\bibnamefont {Tchekhovskoy}},\ and\ \bibinfo {author} {\bibfnamefont {E.}~\bibnamefont {Quataert}},\ }\bibfield  {title} {\bibinfo {title} {{Large-Scale Poloidal Magnetic Field Dynamo Leads to Powerful Jets in GRMHD Simulations of Black Hole Accretion with Toroidal Field}},\ }\href {https://doi.org/10.1093/mnras/staa955} {\bibfield  {journal} {\bibinfo  {journal} {Mon. Not. Roy. Astron. Soc.}\ }\textbf {\bibinfo {volume} {494}},\ \bibinfo {pages} {3656} (\bibinfo {year} {2020})},\ \Eprint {https://arxiv.org/abs/1809.04608} {arXiv:1809.04608 [astro-ph.HE]} \BibitemShut {NoStop}%
\bibitem [{\citenamefont {Hogg}\ and\ \citenamefont {Reynolds}(2018)}]{Hogg:2018zon}%
  \BibitemOpen
  \bibfield  {author} {\bibinfo {author} {\bibfnamefont {J.~D.}\ \bibnamefont {Hogg}}\ and\ \bibinfo {author} {\bibfnamefont {C.}~\bibnamefont {Reynolds}},\ }\bibfield  {title} {\bibinfo {title} {{The influence of accretion disk thickness on the large-scale magnetic dynamo}},\ }\href {https://doi.org/10.3847/1538-4357/aac439} {\bibfield  {journal} {\bibinfo  {journal} {Astrophys. J.}\ }\textbf {\bibinfo {volume} {861}},\ \bibinfo {pages} {24} (\bibinfo {year} {2018})},\ \Eprint {https://arxiv.org/abs/1805.05372} {arXiv:1805.05372 [astro-ph.HE]} \BibitemShut {NoStop}%
\bibitem [{\citenamefont {Jacquemin-Ide}\ \emph {et~al.}(2023)\citenamefont {Jacquemin-Ide}, \citenamefont {Rincon}, \citenamefont {Tchekhovskoy},\ and\ \citenamefont {Liska}}]{Jacquemin-Ide:2023qrj}%
  \BibitemOpen
  \bibfield  {author} {\bibinfo {author} {\bibfnamefont {J.}~\bibnamefont {Jacquemin-Ide}}, \bibinfo {author} {\bibfnamefont {F.}~\bibnamefont {Rincon}}, \bibinfo {author} {\bibfnamefont {A.}~\bibnamefont {Tchekhovskoy}},\ and\ \bibinfo {author} {\bibfnamefont {M.}~\bibnamefont {Liska}},\ }\href@noop {} {\bibinfo {title} {Magnetorotational dynamo can generate large-scale vertical magnetic fields in 3d grmhd simulations of accreting black holes}} (\bibinfo {year} {2023}),\ \Eprint {https://arxiv.org/abs/2311.00034} {arXiv:2311.00034 [astro-ph.HE]} \BibitemShut {NoStop}%
\bibitem [{\citenamefont {Gruzinov}\ and\ \citenamefont {Diamond}(1994)}]{Gruzinov:1994zz}%
  \BibitemOpen
  \bibfield  {author} {\bibinfo {author} {\bibfnamefont {A.~V.}\ \bibnamefont {Gruzinov}}\ and\ \bibinfo {author} {\bibfnamefont {P.~H.}\ \bibnamefont {Diamond}},\ }\bibfield  {title} {\bibinfo {title} {{Self-consistent theory of mean-field electrodynamics}},\ }\href {https://doi.org/10.1103/PhysRevLett.72.1651} {\bibfield  {journal} {\bibinfo  {journal} {Phys. Rev. Lett.}\ }\textbf {\bibinfo {volume} {72}},\ \bibinfo {pages} {1651} (\bibinfo {year} {1994})}\BibitemShut {NoStop}%
\bibitem [{\citenamefont {{Berger}}(1999)}]{1999PPCF...41B.167B}%
  \BibitemOpen
  \bibfield  {author} {\bibinfo {author} {\bibfnamefont {M.~A.}\ \bibnamefont {{Berger}}},\ }\bibfield  {title} {\bibinfo {title} {{Introduction to magnetic helicity.}},\ }\href {https://doi.org/10.1088/0741-3335/41/12B/312} {\bibfield  {journal} {\bibinfo  {journal} {Plasma Physics and Controlled Fusion}\ }\textbf {\bibinfo {volume} {41}},\ \bibinfo {pages} {B167} (\bibinfo {year} {1999})}\BibitemShut {NoStop}%
\bibitem [{\citenamefont {{Taylor}}(1974)}]{1974PhRvL..33.1139T}%
  \BibitemOpen
  \bibfield  {author} {\bibinfo {author} {\bibfnamefont {J.~B.}\ \bibnamefont {{Taylor}}},\ }\bibfield  {title} {\bibinfo {title} {{Relaxation of Toroidal Plasma and Generation of Reverse Magnetic Fields}},\ }\href {https://doi.org/10.1103/PhysRevLett.33.1139} {\bibfield  {journal} {\bibinfo  {journal} {Phys. Rev. Lett.}\ }\textbf {\bibinfo {volume} {33}},\ \bibinfo {pages} {1139} (\bibinfo {year} {1974})}\BibitemShut {NoStop}%
\bibitem [{\citenamefont {{Bhattacharjee}}\ and\ \citenamefont {{Yuan}}(1995)}]{1995ApJ...449..739B}%
  \BibitemOpen
  \bibfield  {author} {\bibinfo {author} {\bibfnamefont {A.}~\bibnamefont {{Bhattacharjee}}}\ and\ \bibinfo {author} {\bibfnamefont {Y.}~\bibnamefont {{Yuan}}},\ }\bibfield  {title} {\bibinfo {title} {{Self-Consistency Constraints on the Dynamo Mechanism}},\ }\href {https://doi.org/10.1086/176094} {\bibfield  {journal} {\bibinfo  {journal} {Astrophys. J.}\ }\textbf {\bibinfo {volume} {449}},\ \bibinfo {pages} {739} (\bibinfo {year} {1995})}\BibitemShut {NoStop}%
\bibitem [{\citenamefont {Blackman}(2015)}]{Blackman:2014kxa}%
  \BibitemOpen
  \bibfield  {author} {\bibinfo {author} {\bibfnamefont {E.~G.}\ \bibnamefont {Blackman}},\ }\bibfield  {title} {\bibinfo {title} {{Magnetic Helicity and Large Scale Magnetic Fields: A Primer}},\ }\href {https://doi.org/10.1007/s11214-014-0038-6} {\bibfield  {journal} {\bibinfo  {journal} {Space Sci. Rev.}\ }\textbf {\bibinfo {volume} {188}},\ \bibinfo {pages} {59} (\bibinfo {year} {2015})},\ \Eprint {https://arxiv.org/abs/1402.0933} {arXiv:1402.0933 [astro-ph.SR]} \BibitemShut {NoStop}%
\bibitem [{\citenamefont {Squire}\ and\ \citenamefont {Bhattacharjee}(2015)}]{Squire:2015jma}%
  \BibitemOpen
  \bibfield  {author} {\bibinfo {author} {\bibfnamefont {J.}~\bibnamefont {Squire}}\ and\ \bibinfo {author} {\bibfnamefont {A.}~\bibnamefont {Bhattacharjee}},\ }\bibfield  {title} {\bibinfo {title} {{Generation of large-scale magnetic fields by small-scale dynamo in shear flows}},\ }\href {https://doi.org/10.1103/PhysRevLett.115.175003} {\bibfield  {journal} {\bibinfo  {journal} {Phys. Rev. Lett.}\ }\textbf {\bibinfo {volume} {115}},\ \bibinfo {pages} {175003} (\bibinfo {year} {2015})},\ \Eprint {https://arxiv.org/abs/1506.04109} {arXiv:1506.04109 [astro-ph.SR]} \BibitemShut {NoStop}%
\bibitem [{\citenamefont {Berger}\ and\ \citenamefont {Field}(1984)}]{Berger_Field_1984}%
  \BibitemOpen
  \bibfield  {author} {\bibinfo {author} {\bibfnamefont {M.~A.}\ \bibnamefont {Berger}}\ and\ \bibinfo {author} {\bibfnamefont {G.~B.}\ \bibnamefont {Field}},\ }\bibfield  {title} {\bibinfo {title} {The topological properties of magnetic helicity},\ }\href {https://doi.org/10.1017/S0022112084002019} {\bibfield  {journal} {\bibinfo  {journal} {Journal of Fluid Mechanics}\ }\textbf {\bibinfo {volume} {147}},\ \bibinfo {pages} {133–148} (\bibinfo {year} {1984})}\BibitemShut {NoStop}%
\bibitem [{\citenamefont {{Moraitis}}\ \emph {et~al.}(2019)\citenamefont {{Moraitis}}, \citenamefont {{Pariat}}, \citenamefont {{Valori}},\ and\ \citenamefont {{Dalmasse}}}]{2019A&A...624A..51M}%
  \BibitemOpen
  \bibfield  {author} {\bibinfo {author} {\bibfnamefont {K.}~\bibnamefont {{Moraitis}}}, \bibinfo {author} {\bibfnamefont {E.}~\bibnamefont {{Pariat}}}, \bibinfo {author} {\bibfnamefont {G.}~\bibnamefont {{Valori}}},\ and\ \bibinfo {author} {\bibfnamefont {K.}~\bibnamefont {{Dalmasse}}},\ }\bibfield  {title} {\bibinfo {title} {{Relative magnetic field line helicity}},\ }\href {https://doi.org/10.1051/0004-6361/201834668} {\bibfield  {journal} {\bibinfo  {journal} {Astron. \& Astrophys.}\ }\textbf {\bibinfo {volume} {624}},\ \bibinfo {eid} {A51} (\bibinfo {year} {2019})},\ \Eprint {https://arxiv.org/abs/1902.10410} {arXiv:1902.10410 [astro-ph.SR]} \BibitemShut {NoStop}%
\bibitem [{\citenamefont {{Brandenburg}}\ \emph {et~al.}(2009)\citenamefont {{Brandenburg}}, \citenamefont {{Candelaresi}},\ and\ \citenamefont {{Chatterjee}}}]{2009MNRAS.398.1414B}%
  \BibitemOpen
  \bibfield  {author} {\bibinfo {author} {\bibfnamefont {A.}~\bibnamefont {{Brandenburg}}}, \bibinfo {author} {\bibfnamefont {S.}~\bibnamefont {{Candelaresi}}},\ and\ \bibinfo {author} {\bibfnamefont {P.}~\bibnamefont {{Chatterjee}}},\ }\bibfield  {title} {\bibinfo {title} {{Small-scale magnetic helicity losses from a mean-field dynamo}},\ }\href {https://doi.org/10.1111/j.1365-2966.2009.15188.x} {\bibfield  {journal} {\bibinfo  {journal} {Mon. Not. Roy. Astron. Soc.}\ }\textbf {\bibinfo {volume} {398}},\ \bibinfo {pages} {1414} (\bibinfo {year} {2009})},\ \Eprint {https://arxiv.org/abs/0905.0242} {arXiv:0905.0242 [astro-ph.SR]} \BibitemShut {NoStop}%
\bibitem [{\citenamefont {Zenati}\ and\ \citenamefont {Vishniac}(2023)}]{Zenati:2021vrr}%
  \BibitemOpen
  \bibfield  {author} {\bibinfo {author} {\bibfnamefont {Y.}~\bibnamefont {Zenati}}\ and\ \bibinfo {author} {\bibfnamefont {E.~T.}\ \bibnamefont {Vishniac}},\ }\bibfield  {title} {\bibinfo {title} {{Conserving Local Magnetic Helicity in Numerical Simulations}},\ }\href {https://doi.org/10.3847/1538-4357/acca1e} {\bibfield  {journal} {\bibinfo  {journal} {Astrophys. J.}\ }\textbf {\bibinfo {volume} {948}},\ \bibinfo {pages} {11} (\bibinfo {year} {2023})},\ \Eprint {https://arxiv.org/abs/2106.06078} {arXiv:2106.06078 [physics.plasm-ph]} \BibitemShut {NoStop}%
\bibitem [{\citenamefont {Schuck}\ and\ \citenamefont {Antiochos}(2019)}]{Schuck_2019}%
  \BibitemOpen
  \bibfield  {author} {\bibinfo {author} {\bibfnamefont {P.~W.}\ \bibnamefont {Schuck}}\ and\ \bibinfo {author} {\bibfnamefont {S.~K.}\ \bibnamefont {Antiochos}},\ }\bibfield  {title} {\bibinfo {title} {Determining the transport of magnetic helicity and free energy in the sun’s atmosphere},\ }\href {https://doi.org/10.3847/1538-4357/ab298a} {\bibfield  {journal} {\bibinfo  {journal} {The Astrophysical Journal}\ }\textbf {\bibinfo {volume} {882}},\ \bibinfo {pages} {151} (\bibinfo {year} {2019})}\BibitemShut {NoStop}%
\bibitem [{\citenamefont {Finn}\ and\ \citenamefont {Antonsen}(1985)}]{Finn_Antonsen_1985}%
  \BibitemOpen
  \bibfield  {author} {\bibinfo {author} {\bibfnamefont {J.}~\bibnamefont {Finn}}\ and\ \bibinfo {author} {\bibfnamefont {T.}~\bibnamefont {Antonsen}},\ }\href@noop {} {\bibfield  {journal} {\bibinfo  {journal} {CoPPC}\ } (\bibinfo {year} {1985})}\BibitemShut {NoStop}%
\bibitem [{\citenamefont {Moffatt}(2014)}]{Moffatt_14}%
  \BibitemOpen
  \bibfield  {author} {\bibinfo {author} {\bibfnamefont {H.~K.}\ \bibnamefont {Moffatt}},\ }\bibfield  {title} {\bibinfo {title} {Helicity and singular structures in fluid dynamics},\ }\href {https://doi.org/10.1073/pnas.1400277111} {\bibfield  {journal} {\bibinfo  {journal} {Proceedings of the National Academy of Sciences}\ }\textbf {\bibinfo {volume} {111}},\ \bibinfo {pages} {3663} (\bibinfo {year} {2014})},\ \Eprint {https://arxiv.org/abs/https://www.pnas.org/doi/pdf/10.1073/pnas.1400277111} {https://www.pnas.org/doi/pdf/10.1073/pnas.1400277111} \BibitemShut {NoStop}%
\bibitem [{\citenamefont {{Valori}}\ \emph {et~al.}(2020)\citenamefont {{Valori}}, \citenamefont {{D{\'e}moulin}}, \citenamefont {{Pariat}}, \citenamefont {{Yeates}}, \citenamefont {{Moraitis}},\ and\ \citenamefont {{Linan}}}]{2020A&A...643A..26V}%
  \BibitemOpen
  \bibfield  {author} {\bibinfo {author} {\bibfnamefont {G.}~\bibnamefont {{Valori}}}, \bibinfo {author} {\bibfnamefont {P.}~\bibnamefont {{D{\'e}moulin}}}, \bibinfo {author} {\bibfnamefont {E.}~\bibnamefont {{Pariat}}}, \bibinfo {author} {\bibfnamefont {A.}~\bibnamefont {{Yeates}}}, \bibinfo {author} {\bibfnamefont {K.}~\bibnamefont {{Moraitis}}},\ and\ \bibinfo {author} {\bibfnamefont {L.}~\bibnamefont {{Linan}}},\ }\bibfield  {title} {\bibinfo {title} {{Additivity of relative magnetic helicity in finite volumes}},\ }\href {https://doi.org/10.1051/0004-6361/202038533} {\bibfield  {journal} {\bibinfo  {journal} {Astron. \& Astrophys.}\ }\textbf {\bibinfo {volume} {643}},\ \bibinfo {eid} {A26} (\bibinfo {year} {2020})},\ \Eprint {https://arxiv.org/abs/2008.00968} {arXiv:2008.00968 [astro-ph.SR]} \BibitemShut {NoStop}%
\bibitem [{\citenamefont {Font}(2008)}]{Font:2008fka}%
  \BibitemOpen
  \bibfield  {author} {\bibinfo {author} {\bibfnamefont {J.~A.}\ \bibnamefont {Font}},\ }\bibfield  {title} {\bibinfo {title} {{Numerical Hydrodynamics and Magnetohydrodynamics in General Relativity}},\ }\href {https://doi.org/10.12942/lrr-2008-7} {\bibfield  {journal} {\bibinfo  {journal} {Living Rev. Rel.}\ }\textbf {\bibinfo {volume} {11}},\ \bibinfo {pages} {7} (\bibinfo {year} {2008})}\BibitemShut {NoStop}%
\bibitem [{\citenamefont {{Bekenstein}}(1987)}]{Bekenstein87}%
  \BibitemOpen
  \bibfield  {author} {\bibinfo {author} {\bibfnamefont {J.~D.}\ \bibnamefont {{Bekenstein}}},\ }\bibfield  {title} {\bibinfo {title} {{Helicity Conservation Laws for Fluids and Plasmas}},\ }\href {https://doi.org/10.1086/165447} {\bibfield  {journal} {\bibinfo  {journal} {Astrophys. J.}\ }\textbf {\bibinfo {volume} {319}},\ \bibinfo {pages} {207} (\bibinfo {year} {1987})}\BibitemShut {NoStop}%
\bibitem [{\citenamefont {Baumgarte}\ and\ \citenamefont {Shapiro}(2003)}]{Baumgarte:2002vv}%
  \BibitemOpen
  \bibfield  {author} {\bibinfo {author} {\bibfnamefont {T.~W.}\ \bibnamefont {Baumgarte}}\ and\ \bibinfo {author} {\bibfnamefont {S.~L.}\ \bibnamefont {Shapiro}},\ }\bibfield  {title} {\bibinfo {title} {{General - relativistic MHD for the numerical construction of dynamical space - times}},\ }\href {https://doi.org/10.1086/346103} {\bibfield  {journal} {\bibinfo  {journal} {Astrophys. J.}\ }\textbf {\bibinfo {volume} {585}},\ \bibinfo {pages} {921} (\bibinfo {year} {2003})},\ \Eprint {https://arxiv.org/abs/astro-ph/0211340} {arXiv:astro-ph/0211340} \BibitemShut {NoStop}%
\bibitem [{\citenamefont {Arnowitt}\ \emph {et~al.}(2008)\citenamefont {Arnowitt}, \citenamefont {Deser},\ and\ \citenamefont {Misner}}]{Arnowitt:1962hi}%
  \BibitemOpen
  \bibfield  {author} {\bibinfo {author} {\bibfnamefont {R.~L.}\ \bibnamefont {Arnowitt}}, \bibinfo {author} {\bibfnamefont {S.}~\bibnamefont {Deser}},\ and\ \bibinfo {author} {\bibfnamefont {C.~W.}\ \bibnamefont {Misner}},\ }\bibfield  {title} {\bibinfo {title} {{The Dynamics of general relativity}},\ }\href {https://doi.org/10.1007/s10714-008-0661-1} {\bibfield  {journal} {\bibinfo  {journal} {Gen. Rel. Grav.}\ }\textbf {\bibinfo {volume} {40}},\ \bibinfo {pages} {1997} (\bibinfo {year} {2008})},\ \Eprint {https://arxiv.org/abs/gr-qc/0405109} {arXiv:gr-qc/0405109} \BibitemShut {NoStop}%
\bibitem [{\citenamefont {Komissarov}(2004)}]{Komissarov:2004ms}%
  \BibitemOpen
  \bibfield  {author} {\bibinfo {author} {\bibfnamefont {S.~S.}\ \bibnamefont {Komissarov}},\ }\bibfield  {title} {\bibinfo {title} {{Electrodynamics of black hole magnetospheres}},\ }\href {https://doi.org/10.1111/j.1365-2966.2004.07446.x} {\bibfield  {journal} {\bibinfo  {journal} {Mon. Not. Roy. Astron. Soc.}\ }\textbf {\bibinfo {volume} {350}},\ \bibinfo {pages} {407} (\bibinfo {year} {2004})},\ \Eprint {https://arxiv.org/abs/astro-ph/0402403} {arXiv:astro-ph/0402403} \BibitemShut {NoStop}%
\bibitem [{\citenamefont {{Russell}}\ \emph {et~al.}(2019)\citenamefont {{Russell}}, \citenamefont {{Demoulin}}, \citenamefont {{Hornig}}, \citenamefont {{Pontin}},\ and\ \citenamefont {{Candelaresi}}}]{2019ApJ...884...55R}%
  \BibitemOpen
  \bibfield  {author} {\bibinfo {author} {\bibfnamefont {A.~J.~B.}\ \bibnamefont {{Russell}}}, \bibinfo {author} {\bibfnamefont {P.}~\bibnamefont {{Demoulin}}}, \bibinfo {author} {\bibfnamefont {G.}~\bibnamefont {{Hornig}}}, \bibinfo {author} {\bibfnamefont {D.~I.}\ \bibnamefont {{Pontin}}},\ and\ \bibinfo {author} {\bibfnamefont {S.}~\bibnamefont {{Candelaresi}}},\ }\bibfield  {title} {\bibinfo {title} {{Do Current and Magnetic Helicities Have the Same Sign?}},\ }\href {https://doi.org/10.3847/1538-4357/ab40b4} {\bibfield  {journal} {\bibinfo  {journal} {Astrophys. J.}\ }\textbf {\bibinfo {volume} {884}},\ \bibinfo {eid} {55} (\bibinfo {year} {2019})}\BibitemShut {NoStop}%
\bibitem [{\citenamefont {Backus}(1986)}]{Backus_1986}%
  \BibitemOpen
  \bibfield  {author} {\bibinfo {author} {\bibfnamefont {G.}~\bibnamefont {Backus}},\ }\bibfield  {title} {\bibinfo {title} {Poloidal and toroidal fields in geomagnetic field modeling},\ }\href {https://doi.org/https://doi.org/10.1029/RG024i001p00075} {\bibfield  {journal} {\bibinfo  {journal} {Reviews of Geophysics}\ }\textbf {\bibinfo {volume} {24}},\ \bibinfo {pages} {75} (\bibinfo {year} {1986})}\BibitemShut {NoStop}%
\bibitem [{\citenamefont {Baiotti}\ and\ \citenamefont {Rezzolla}(2017)}]{Baiotti:2016qnr}%
  \BibitemOpen
  \bibfield  {author} {\bibinfo {author} {\bibfnamefont {L.}~\bibnamefont {Baiotti}}\ and\ \bibinfo {author} {\bibfnamefont {L.}~\bibnamefont {Rezzolla}},\ }\bibfield  {title} {\bibinfo {title} {{Binary neutron star mergers: a review of Einstein\textquoteright{}s richest laboratory}},\ }\href {https://doi.org/10.1088/1361-6633/aa67bb} {\bibfield  {journal} {\bibinfo  {journal} {Rept. Prog. Phys.}\ }\textbf {\bibinfo {volume} {80}},\ \bibinfo {pages} {096901} (\bibinfo {year} {2017})},\ \Eprint {https://arxiv.org/abs/1607.03540} {arXiv:1607.03540 [gr-qc]} \BibitemShut {NoStop}%
\bibitem [{\citenamefont {Radice}\ \emph {et~al.}(2020)\citenamefont {Radice}, \citenamefont {Bernuzzi},\ and\ \citenamefont {Perego}}]{Radice:2020ddv}%
  \BibitemOpen
  \bibfield  {author} {\bibinfo {author} {\bibfnamefont {D.}~\bibnamefont {Radice}}, \bibinfo {author} {\bibfnamefont {S.}~\bibnamefont {Bernuzzi}},\ and\ \bibinfo {author} {\bibfnamefont {A.}~\bibnamefont {Perego}},\ }\bibfield  {title} {\bibinfo {title} {{The Dynamics of Binary Neutron Star Mergers and GW170817}},\ }\href {https://doi.org/10.1146/annurev-nucl-013120-114541} {\bibfield  {journal} {\bibinfo  {journal} {Ann. Rev. Nucl. Part. Sci.}\ }\textbf {\bibinfo {volume} {70}},\ \bibinfo {pages} {95} (\bibinfo {year} {2020})},\ \Eprint {https://arxiv.org/abs/2002.03863} {arXiv:2002.03863 [astro-ph.HE]} \BibitemShut {NoStop}%
\bibitem [{\citenamefont {Kastaun}\ \emph {et~al.}(2016)\citenamefont {Kastaun}, \citenamefont {Ciolfi},\ and\ \citenamefont {Giacomazzo}}]{Kastaun:2016yaf}%
  \BibitemOpen
  \bibfield  {author} {\bibinfo {author} {\bibfnamefont {W.}~\bibnamefont {Kastaun}}, \bibinfo {author} {\bibfnamefont {R.}~\bibnamefont {Ciolfi}},\ and\ \bibinfo {author} {\bibfnamefont {B.}~\bibnamefont {Giacomazzo}},\ }\bibfield  {title} {\bibinfo {title} {{Structure of Stable Binary Neutron Star Merger Remnants: a Case Study}},\ }\href {https://doi.org/10.1103/PhysRevD.94.044060} {\bibfield  {journal} {\bibinfo  {journal} {Phys. Rev. D}\ }\textbf {\bibinfo {volume} {94}},\ \bibinfo {pages} {044060} (\bibinfo {year} {2016})},\ \Eprint {https://arxiv.org/abs/1607.02186} {arXiv:1607.02186 [astro-ph.HE]} \BibitemShut {NoStop}%
\bibitem [{\citenamefont {Kiuchi}\ \emph {et~al.}(2015)\citenamefont {Kiuchi}, \citenamefont {Cerd\'a-Dur\'an}, \citenamefont {Kyutoku}, \citenamefont {Sekiguchi},\ and\ \citenamefont {Shibata}}]{Kiuchi:2015sga}%
  \BibitemOpen
  \bibfield  {author} {\bibinfo {author} {\bibfnamefont {K.}~\bibnamefont {Kiuchi}}, \bibinfo {author} {\bibfnamefont {P.}~\bibnamefont {Cerd\'a-Dur\'an}}, \bibinfo {author} {\bibfnamefont {K.}~\bibnamefont {Kyutoku}}, \bibinfo {author} {\bibfnamefont {Y.}~\bibnamefont {Sekiguchi}},\ and\ \bibinfo {author} {\bibfnamefont {M.}~\bibnamefont {Shibata}},\ }\bibfield  {title} {\bibinfo {title} {{Efficient magnetic-field amplification due to the Kelvin-Helmholtz instability in binary neutron star mergers}},\ }\href {https://doi.org/10.1103/PhysRevD.92.124034} {\bibfield  {journal} {\bibinfo  {journal} {Phys. Rev. D}\ }\textbf {\bibinfo {volume} {92}},\ \bibinfo {pages} {124034} (\bibinfo {year} {2015})},\ \Eprint {https://arxiv.org/abs/1509.09205} {arXiv:1509.09205 [astro-ph.HE]} \BibitemShut {NoStop}%
\bibitem [{\citenamefont {Most}(2023)}]{Most:2023sme}%
  \BibitemOpen
  \bibfield  {author} {\bibinfo {author} {\bibfnamefont {E.~R.}\ \bibnamefont {Most}},\ }\bibfield  {title} {\bibinfo {title} {{Impact of a mean field dynamo on neutron star mergers leading to magnetar remnants}},\ }\href {https://doi.org/10.1103/PhysRevD.108.123012} {\bibfield  {journal} {\bibinfo  {journal} {Phys. Rev. D}\ }\textbf {\bibinfo {volume} {108}},\ \bibinfo {pages} {123012} (\bibinfo {year} {2023})},\ \Eprint {https://arxiv.org/abs/2311.03333} {arXiv:2311.03333 [astro-ph.HE]} \BibitemShut {NoStop}%
\bibitem [{\citenamefont {Palenzuela}\ \emph {et~al.}(2015)\citenamefont {Palenzuela}, \citenamefont {Liebling}, \citenamefont {Neilsen}, \citenamefont {Lehner}, \citenamefont {Caballero}, \citenamefont {O'Connor},\ and\ \citenamefont {Anderson}}]{Palenzuela:2015dqa}%
  \BibitemOpen
  \bibfield  {author} {\bibinfo {author} {\bibfnamefont {C.}~\bibnamefont {Palenzuela}}, \bibinfo {author} {\bibfnamefont {S.~L.}\ \bibnamefont {Liebling}}, \bibinfo {author} {\bibfnamefont {D.}~\bibnamefont {Neilsen}}, \bibinfo {author} {\bibfnamefont {L.}~\bibnamefont {Lehner}}, \bibinfo {author} {\bibfnamefont {O.~L.}\ \bibnamefont {Caballero}}, \bibinfo {author} {\bibfnamefont {E.}~\bibnamefont {O'Connor}},\ and\ \bibinfo {author} {\bibfnamefont {M.}~\bibnamefont {Anderson}},\ }\bibfield  {title} {\bibinfo {title} {{Effects of the microphysical Equation of State in the mergers of magnetized Neutron Stars With Neutrino Cooling}},\ }\href {https://doi.org/10.1103/PhysRevD.92.044045} {\bibfield  {journal} {\bibinfo  {journal} {Phys. Rev. D}\ }\textbf {\bibinfo {volume} {92}},\ \bibinfo {pages} {044045} (\bibinfo {year} {2015})},\ \Eprint {https://arxiv.org/abs/1505.01607} {arXiv:1505.01607 [gr-qc]} \BibitemShut {NoStop}%
\bibitem [{\citenamefont {Ciolfi}\ \emph {et~al.}(2019)\citenamefont {Ciolfi}, \citenamefont {Kastaun}, \citenamefont {Kalinani},\ and\ \citenamefont {Giacomazzo}}]{Ciolfi:2019fie}%
  \BibitemOpen
  \bibfield  {author} {\bibinfo {author} {\bibfnamefont {R.}~\bibnamefont {Ciolfi}}, \bibinfo {author} {\bibfnamefont {W.}~\bibnamefont {Kastaun}}, \bibinfo {author} {\bibfnamefont {J.~V.}\ \bibnamefont {Kalinani}},\ and\ \bibinfo {author} {\bibfnamefont {B.}~\bibnamefont {Giacomazzo}},\ }\bibfield  {title} {\bibinfo {title} {{First 100 ms of a long-lived magnetized neutron star formed in a binary neutron star merger}},\ }\href {https://doi.org/10.1103/PhysRevD.100.023005} {\bibfield  {journal} {\bibinfo  {journal} {Phys. Rev. D}\ }\textbf {\bibinfo {volume} {100}},\ \bibinfo {pages} {023005} (\bibinfo {year} {2019})},\ \Eprint {https://arxiv.org/abs/1904.10222} {arXiv:1904.10222 [astro-ph.HE]} \BibitemShut {NoStop}%
\bibitem [{\citenamefont {Ruiz}\ \emph {et~al.}(2016)\citenamefont {Ruiz}, \citenamefont {Lang}, \citenamefont {Paschalidis},\ and\ \citenamefont {Shapiro}}]{Ruiz:2016rai}%
  \BibitemOpen
  \bibfield  {author} {\bibinfo {author} {\bibfnamefont {M.}~\bibnamefont {Ruiz}}, \bibinfo {author} {\bibfnamefont {R.~N.}\ \bibnamefont {Lang}}, \bibinfo {author} {\bibfnamefont {V.}~\bibnamefont {Paschalidis}},\ and\ \bibinfo {author} {\bibfnamefont {S.~L.}\ \bibnamefont {Shapiro}},\ }\bibfield  {title} {\bibinfo {title} {{Binary Neutron Star Mergers: a jet Engine for Short Gamma-ray Bursts}},\ }\href {https://doi.org/10.3847/2041-8205/824/1/L6} {\bibfield  {journal} {\bibinfo  {journal} {Astrophys. J. Lett.}\ }\textbf {\bibinfo {volume} {824}},\ \bibinfo {pages} {L6} (\bibinfo {year} {2016})},\ \Eprint {https://arxiv.org/abs/1604.02455} {arXiv:1604.02455 [astro-ph.HE]} \BibitemShut {NoStop}%
\bibitem [{\citenamefont {Aguilera-Miret}\ \emph {et~al.}(2022)\citenamefont {Aguilera-Miret}, \citenamefont {Vigan\`o},\ and\ \citenamefont {Palenzuela}}]{Aguilera-Miret:2021fre}%
  \BibitemOpen
  \bibfield  {author} {\bibinfo {author} {\bibfnamefont {R.}~\bibnamefont {Aguilera-Miret}}, \bibinfo {author} {\bibfnamefont {D.}~\bibnamefont {Vigan\`o}},\ and\ \bibinfo {author} {\bibfnamefont {C.}~\bibnamefont {Palenzuela}},\ }\bibfield  {title} {\bibinfo {title} {{Universality of the Turbulent Magnetic Field in Hypermassive Neutron Stars Produced by Binary Mergers}},\ }\href {https://doi.org/10.3847/2041-8213/ac50a7} {\bibfield  {journal} {\bibinfo  {journal} {Astrophys. J. Lett.}\ }\textbf {\bibinfo {volume} {926}},\ \bibinfo {pages} {L31} (\bibinfo {year} {2022})},\ \Eprint {https://arxiv.org/abs/2112.08406} {arXiv:2112.08406 [gr-qc]} \BibitemShut {NoStop}%
\bibitem [{\citenamefont {Chabanov}\ \emph {et~al.}(2023)\citenamefont {Chabanov}, \citenamefont {Tootle}, \citenamefont {Most},\ and\ \citenamefont {Rezzolla}}]{Chabanov:2022twz}%
  \BibitemOpen
  \bibfield  {author} {\bibinfo {author} {\bibfnamefont {M.}~\bibnamefont {Chabanov}}, \bibinfo {author} {\bibfnamefont {S.~D.}\ \bibnamefont {Tootle}}, \bibinfo {author} {\bibfnamefont {E.~R.}\ \bibnamefont {Most}},\ and\ \bibinfo {author} {\bibfnamefont {L.}~\bibnamefont {Rezzolla}},\ }\bibfield  {title} {\bibinfo {title} {{Crustal Magnetic Fields Do Not Lead to Large Magnetic-field Amplifications in Binary Neutron Star Mergers}},\ }\href {https://doi.org/10.3847/2041-8213/acbbc5} {\bibfield  {journal} {\bibinfo  {journal} {Astrophys. J. Lett.}\ }\textbf {\bibinfo {volume} {945}},\ \bibinfo {pages} {L14} (\bibinfo {year} {2023})},\ \Eprint {https://arxiv.org/abs/2211.13661} {arXiv:2211.13661 [astro-ph.HE]} \BibitemShut {NoStop}%
\bibitem [{\citenamefont {Shapiro}(2000)}]{Shapiro:2000zh}%
  \BibitemOpen
  \bibfield  {author} {\bibinfo {author} {\bibfnamefont {S.~L.}\ \bibnamefont {Shapiro}},\ }\bibfield  {title} {\bibinfo {title} {{Differential rotation in neutron stars: Magnetic braking and viscous damping}},\ }\href {https://doi.org/10.1086/317209} {\bibfield  {journal} {\bibinfo  {journal} {Astrophys. J.}\ }\textbf {\bibinfo {volume} {544}},\ \bibinfo {pages} {397} (\bibinfo {year} {2000})},\ \Eprint {https://arxiv.org/abs/astro-ph/0010493} {arXiv:astro-ph/0010493} \BibitemShut {NoStop}%
\bibitem [{\citenamefont {Most}\ \emph {et~al.}(2023)\citenamefont {Most}, \citenamefont {Motornenko}, \citenamefont {Steinheimer}, \citenamefont {Dexheimer}, \citenamefont {Hanauske}, \citenamefont {Rezzolla},\ and\ \citenamefont {Stoecker}}]{Most:2022wgo}%
  \BibitemOpen
  \bibfield  {author} {\bibinfo {author} {\bibfnamefont {E.~R.}\ \bibnamefont {Most}}, \bibinfo {author} {\bibfnamefont {A.}~\bibnamefont {Motornenko}}, \bibinfo {author} {\bibfnamefont {J.}~\bibnamefont {Steinheimer}}, \bibinfo {author} {\bibfnamefont {V.}~\bibnamefont {Dexheimer}}, \bibinfo {author} {\bibfnamefont {M.}~\bibnamefont {Hanauske}}, \bibinfo {author} {\bibfnamefont {L.}~\bibnamefont {Rezzolla}},\ and\ \bibinfo {author} {\bibfnamefont {H.}~\bibnamefont {Stoecker}},\ }\bibfield  {title} {\bibinfo {title} {{Probing neutron-star matter in the lab: Similarities and differences between binary mergers and heavy-ion collisions}},\ }\href {https://doi.org/10.1103/PhysRevD.107.043034} {\bibfield  {journal} {\bibinfo  {journal} {Phys. Rev. D}\ }\textbf {\bibinfo {volume} {107}},\ \bibinfo {pages} {043034} (\bibinfo {year} {2023})},\ \Eprint {https://arxiv.org/abs/2201.13150} {arXiv:2201.13150 [nucl-th]} \BibitemShut {NoStop}%
\bibitem [{\citenamefont {Hodge}\ and\ \citenamefont {Hodge}(1989)}]{hodge1989theory}%
  \BibitemOpen
  \bibfield  {author} {\bibinfo {author} {\bibfnamefont {W.~V.~D.}\ \bibnamefont {Hodge}}\ and\ \bibinfo {author} {\bibfnamefont {W.~V.~D.}\ \bibnamefont {Hodge}},\ }\href@noop {} {\emph {\bibinfo {title} {The theory and applications of harmonic integrals}}}\ (\bibinfo  {publisher} {CUP Archive},\ \bibinfo {year} {1989})\BibitemShut {NoStop}%
\bibitem [{\citenamefont {Bernuzzi}\ and\ \citenamefont {Hilditch}(2010)}]{Bernuzzi:2009ex}%
  \BibitemOpen
  \bibfield  {author} {\bibinfo {author} {\bibfnamefont {S.}~\bibnamefont {Bernuzzi}}\ and\ \bibinfo {author} {\bibfnamefont {D.}~\bibnamefont {Hilditch}},\ }\bibfield  {title} {\bibinfo {title} {{Constraint violation in free evolution schemes: Comparing BSSNOK with a conformal decomposition of Z4}},\ }\href {https://doi.org/10.1103/PhysRevD.81.084003} {\bibfield  {journal} {\bibinfo  {journal} {Phys. Rev. D}\ }\textbf {\bibinfo {volume} {81}},\ \bibinfo {pages} {084003} (\bibinfo {year} {2010})},\ \Eprint {https://arxiv.org/abs/0912.2920} {arXiv:0912.2920 [gr-qc]} \BibitemShut {NoStop}%
\bibitem [{\citenamefont {Hilditch}\ \emph {et~al.}(2013)\citenamefont {Hilditch}, \citenamefont {Bernuzzi}, \citenamefont {Thierfelder}, \citenamefont {Cao}, \citenamefont {Tichy},\ and\ \citenamefont {Bruegmann}}]{Hilditch:2012fp}%
  \BibitemOpen
  \bibfield  {author} {\bibinfo {author} {\bibfnamefont {D.}~\bibnamefont {Hilditch}}, \bibinfo {author} {\bibfnamefont {S.}~\bibnamefont {Bernuzzi}}, \bibinfo {author} {\bibfnamefont {M.}~\bibnamefont {Thierfelder}}, \bibinfo {author} {\bibfnamefont {Z.}~\bibnamefont {Cao}}, \bibinfo {author} {\bibfnamefont {W.}~\bibnamefont {Tichy}},\ and\ \bibinfo {author} {\bibfnamefont {B.}~\bibnamefont {Bruegmann}},\ }\bibfield  {title} {\bibinfo {title} {{Compact binary evolutions with the Z4c formulation}},\ }\href {https://doi.org/10.1103/PhysRevD.88.084057} {\bibfield  {journal} {\bibinfo  {journal} {Phys. Rev. D}\ }\textbf {\bibinfo {volume} {88}},\ \bibinfo {pages} {084057} (\bibinfo {year} {2013})},\ \Eprint {https://arxiv.org/abs/1212.2901} {arXiv:1212.2901 [gr-qc]} \BibitemShut {NoStop}%
\bibitem [{\citenamefont {Alcubierre}\ \emph {et~al.}(2003)\citenamefont {Alcubierre}, \citenamefont {Bruegmann}, \citenamefont {Diener}, \citenamefont {Koppitz}, \citenamefont {Pollney}, \citenamefont {Seidel},\ and\ \citenamefont {Takahashi}}]{Alcubierre:2002kk}%
  \BibitemOpen
  \bibfield  {author} {\bibinfo {author} {\bibfnamefont {M.}~\bibnamefont {Alcubierre}}, \bibinfo {author} {\bibfnamefont {B.}~\bibnamefont {Bruegmann}}, \bibinfo {author} {\bibfnamefont {P.}~\bibnamefont {Diener}}, \bibinfo {author} {\bibfnamefont {M.}~\bibnamefont {Koppitz}}, \bibinfo {author} {\bibfnamefont {D.}~\bibnamefont {Pollney}}, \bibinfo {author} {\bibfnamefont {E.}~\bibnamefont {Seidel}},\ and\ \bibinfo {author} {\bibfnamefont {R.}~\bibnamefont {Takahashi}},\ }\bibfield  {title} {\bibinfo {title} {{Gauge conditions for long term numerical black hole evolutions without excision}},\ }\href {https://doi.org/10.1103/PhysRevD.67.084023} {\bibfield  {journal} {\bibinfo  {journal} {Phys. Rev. D}\ }\textbf {\bibinfo {volume} {67}},\ \bibinfo {pages} {084023} (\bibinfo {year} {2003})},\ \Eprint {https://arxiv.org/abs/gr-qc/0206072} {arXiv:gr-qc/0206072} \BibitemShut {NoStop}%
\bibitem [{\citenamefont {Duez}\ \emph {et~al.}(2005)\citenamefont {Duez}, \citenamefont {Liu}, \citenamefont {Shapiro},\ and\ \citenamefont {Stephens}}]{Duez:2005sf}%
  \BibitemOpen
  \bibfield  {author} {\bibinfo {author} {\bibfnamefont {M.~D.}\ \bibnamefont {Duez}}, \bibinfo {author} {\bibfnamefont {Y.~T.}\ \bibnamefont {Liu}}, \bibinfo {author} {\bibfnamefont {S.~L.}\ \bibnamefont {Shapiro}},\ and\ \bibinfo {author} {\bibfnamefont {B.~C.}\ \bibnamefont {Stephens}},\ }\bibfield  {title} {\bibinfo {title} {{Relativistic magnetohydrodynamics in dynamical spacetimes: Numerical methods and tests}},\ }\href {https://doi.org/10.1103/PhysRevD.72.024028} {\bibfield  {journal} {\bibinfo  {journal} {Phys. Rev. D}\ }\textbf {\bibinfo {volume} {72}},\ \bibinfo {pages} {024028} (\bibinfo {year} {2005})},\ \Eprint {https://arxiv.org/abs/astro-ph/0503420} {arXiv:astro-ph/0503420} \BibitemShut {NoStop}%
\bibitem [{\citenamefont {Etienne}\ \emph {et~al.}(2010)\citenamefont {Etienne}, \citenamefont {Liu},\ and\ \citenamefont {Shapiro}}]{Etienne:2010ui}%
  \BibitemOpen
  \bibfield  {author} {\bibinfo {author} {\bibfnamefont {Z.~B.}\ \bibnamefont {Etienne}}, \bibinfo {author} {\bibfnamefont {Y.~T.}\ \bibnamefont {Liu}},\ and\ \bibinfo {author} {\bibfnamefont {S.~L.}\ \bibnamefont {Shapiro}},\ }\bibfield  {title} {\bibinfo {title} {{Relativistic magnetohydrodynamics in dynamical spacetimes: A new AMR implementation}},\ }\href {https://doi.org/10.1103/PhysRevD.82.084031} {\bibfield  {journal} {\bibinfo  {journal} {Phys. Rev. D}\ }\textbf {\bibinfo {volume} {82}},\ \bibinfo {pages} {084031} (\bibinfo {year} {2010})},\ \Eprint {https://arxiv.org/abs/1007.2848} {arXiv:1007.2848 [astro-ph.HE]} \BibitemShut {NoStop}%
\bibitem [{\citenamefont {Etienne}\ \emph {et~al.}(2012)\citenamefont {Etienne}, \citenamefont {Paschalidis}, \citenamefont {Liu},\ and\ \citenamefont {Shapiro}}]{Etienne:2011re}%
  \BibitemOpen
  \bibfield  {author} {\bibinfo {author} {\bibfnamefont {Z.~B.}\ \bibnamefont {Etienne}}, \bibinfo {author} {\bibfnamefont {V.}~\bibnamefont {Paschalidis}}, \bibinfo {author} {\bibfnamefont {Y.~T.}\ \bibnamefont {Liu}},\ and\ \bibinfo {author} {\bibfnamefont {S.~L.}\ \bibnamefont {Shapiro}},\ }\bibfield  {title} {\bibinfo {title} {{Relativistic MHD in dynamical spacetimes: Improved EM gauge condition for AMR grids}},\ }\href {https://doi.org/10.1103/PhysRevD.85.024013} {\bibfield  {journal} {\bibinfo  {journal} {Phys. Rev. D}\ }\textbf {\bibinfo {volume} {85}},\ \bibinfo {pages} {024013} (\bibinfo {year} {2012})},\ \Eprint {https://arxiv.org/abs/1110.4633} {arXiv:1110.4633 [astro-ph.HE]} \BibitemShut {NoStop}%
\bibitem [{\citenamefont {Most}\ \emph {et~al.}(2019)\citenamefont {Most}, \citenamefont {Papenfort},\ and\ \citenamefont {Rezzolla}}]{Most:2019kfe}%
  \BibitemOpen
  \bibfield  {author} {\bibinfo {author} {\bibfnamefont {E.~R.}\ \bibnamefont {Most}}, \bibinfo {author} {\bibfnamefont {L.~J.}\ \bibnamefont {Papenfort}},\ and\ \bibinfo {author} {\bibfnamefont {L.}~\bibnamefont {Rezzolla}},\ }\bibfield  {title} {\bibinfo {title} {{Beyond second-order convergence in simulations of magnetized binary neutron stars with realistic microphysics}},\ }\href {https://doi.org/10.1093/mnras/stz2809} {\bibfield  {journal} {\bibinfo  {journal} {Mon. Not. Roy. Astron. Soc.}\ }\textbf {\bibinfo {volume} {490}},\ \bibinfo {pages} {3588} (\bibinfo {year} {2019})},\ \Eprint {https://arxiv.org/abs/1907.10328} {arXiv:1907.10328 [astro-ph.HE]} \BibitemShut {NoStop}%
\bibitem [{\citenamefont {Etienne}\ \emph {et~al.}(2015)\citenamefont {Etienne}, \citenamefont {Paschalidis}, \citenamefont {Haas}, \citenamefont {M\"osta},\ and\ \citenamefont {Shapiro}}]{Etienne:2015cea}%
  \BibitemOpen
  \bibfield  {author} {\bibinfo {author} {\bibfnamefont {Z.~B.}\ \bibnamefont {Etienne}}, \bibinfo {author} {\bibfnamefont {V.}~\bibnamefont {Paschalidis}}, \bibinfo {author} {\bibfnamefont {R.}~\bibnamefont {Haas}}, \bibinfo {author} {\bibfnamefont {P.}~\bibnamefont {M\"osta}},\ and\ \bibinfo {author} {\bibfnamefont {S.~L.}\ \bibnamefont {Shapiro}},\ }\bibfield  {title} {\bibinfo {title} {{IllinoisGRMHD: An Open-Source, User-Friendly GRMHD Code for Dynamical Spacetimes}},\ }\href {https://doi.org/10.1088/0264-9381/32/17/175009} {\bibfield  {journal} {\bibinfo  {journal} {Class. Quant. Grav.}\ }\textbf {\bibinfo {volume} {32}},\ \bibinfo {pages} {175009} (\bibinfo {year} {2015})},\ \Eprint {https://arxiv.org/abs/1501.07276} {arXiv:1501.07276 [astro-ph.HE]} \BibitemShut {NoStop}%
\bibitem [{\citenamefont {Loffler}\ \emph {et~al.}(2012)\citenamefont {Loffler} \emph {et~al.}}]{Loffler:2011ay}%
  \BibitemOpen
  \bibfield  {author} {\bibinfo {author} {\bibfnamefont {F.}~\bibnamefont {Loffler}} \emph {et~al.},\ }\bibfield  {title} {\bibinfo {title} {{The Einstein Toolkit: A Community Computational Infrastructure for Relativistic Astrophysics}},\ }\href {https://doi.org/10.1088/0264-9381/29/11/115001} {\bibfield  {journal} {\bibinfo  {journal} {Class. Quant. Grav.}\ }\textbf {\bibinfo {volume} {29}},\ \bibinfo {pages} {115001} (\bibinfo {year} {2012})},\ \Eprint {https://arxiv.org/abs/1111.3344} {arXiv:1111.3344 [gr-qc]} \BibitemShut {NoStop}%
\bibitem [{\citenamefont {Zlochower}\ \emph {et~al.}(2005)\citenamefont {Zlochower}, \citenamefont {Baker}, \citenamefont {Campanelli},\ and\ \citenamefont {Lousto}}]{Zlochower:2005bj}%
  \BibitemOpen
  \bibfield  {author} {\bibinfo {author} {\bibfnamefont {Y.}~\bibnamefont {Zlochower}}, \bibinfo {author} {\bibfnamefont {J.~G.}\ \bibnamefont {Baker}}, \bibinfo {author} {\bibfnamefont {M.}~\bibnamefont {Campanelli}},\ and\ \bibinfo {author} {\bibfnamefont {C.~O.}\ \bibnamefont {Lousto}},\ }\bibfield  {title} {\bibinfo {title} {{Accurate black hole evolutions by fourth-order numerical relativity}},\ }\href {https://doi.org/10.1103/PhysRevD.72.024021} {\bibfield  {journal} {\bibinfo  {journal} {Phys. Rev. D}\ }\textbf {\bibinfo {volume} {72}},\ \bibinfo {pages} {024021} (\bibinfo {year} {2005})},\ \Eprint {https://arxiv.org/abs/gr-qc/0505055} {arXiv:gr-qc/0505055} \BibitemShut {NoStop}%
\bibitem [{\citenamefont {Del~Zanna}\ \emph {et~al.}(2007)\citenamefont {Del~Zanna}, \citenamefont {Zanotti}, \citenamefont {Bucciantini},\ and\ \citenamefont {Londrillo}}]{DelZanna:2007pk}%
  \BibitemOpen
  \bibfield  {author} {\bibinfo {author} {\bibfnamefont {L.}~\bibnamefont {Del~Zanna}}, \bibinfo {author} {\bibfnamefont {O.}~\bibnamefont {Zanotti}}, \bibinfo {author} {\bibfnamefont {N.}~\bibnamefont {Bucciantini}},\ and\ \bibinfo {author} {\bibfnamefont {P.}~\bibnamefont {Londrillo}},\ }\bibfield  {title} {\bibinfo {title} {{ECHO: an Eulerian Conservative High Order scheme for general relativistic magnetohydrodynamics and magnetodynamics}},\ }\href {https://doi.org/10.1051/0004-6361:20077093} {\bibfield  {journal} {\bibinfo  {journal} {Astron. Astrophys.}\ }\textbf {\bibinfo {volume} {473}},\ \bibinfo {pages} {11} (\bibinfo {year} {2007})},\ \Eprint {https://arxiv.org/abs/0704.3206} {arXiv:0704.3206 [astro-ph]} \BibitemShut {NoStop}%
\bibitem [{\citenamefont {Hempel}\ and\ \citenamefont {Schaffner-Bielich}(2010)}]{Hempel:2009mc}%
  \BibitemOpen
  \bibfield  {author} {\bibinfo {author} {\bibfnamefont {M.}~\bibnamefont {Hempel}}\ and\ \bibinfo {author} {\bibfnamefont {J.}~\bibnamefont {Schaffner-Bielich}},\ }\bibfield  {title} {\bibinfo {title} {{Statistical Model for a Complete Supernova Equation of State}},\ }\href {https://doi.org/10.1016/j.nuclphysa.2010.02.010} {\bibfield  {journal} {\bibinfo  {journal} {Nucl. Phys. A}\ }\textbf {\bibinfo {volume} {837}},\ \bibinfo {pages} {210} (\bibinfo {year} {2010})},\ \Eprint {https://arxiv.org/abs/0911.4073} {arXiv:0911.4073 [nucl-th]} \BibitemShut {NoStop}%
\bibitem [{\citenamefont {Grandclement}(2010)}]{Grandclement:2009ju}%
  \BibitemOpen
  \bibfield  {author} {\bibinfo {author} {\bibfnamefont {P.}~\bibnamefont {Grandclement}},\ }\bibfield  {title} {\bibinfo {title} {{Kadath: A Spectral solver for theoretical physics}},\ }\href {https://doi.org/10.1016/j.jcp.2010.01.005} {\bibfield  {journal} {\bibinfo  {journal} {J. Comput. Phys.}\ }\textbf {\bibinfo {volume} {229}},\ \bibinfo {pages} {3334} (\bibinfo {year} {2010})},\ \Eprint {https://arxiv.org/abs/0909.1228} {arXiv:0909.1228 [gr-qc]} \BibitemShut {NoStop}%
\bibitem [{\citenamefont {Papenfort}\ \emph {et~al.}(2021)\citenamefont {Papenfort}, \citenamefont {Tootle}, \citenamefont {Grandcl\'ement}, \citenamefont {Most},\ and\ \citenamefont {Rezzolla}}]{Papenfort:2021hod}%
  \BibitemOpen
  \bibfield  {author} {\bibinfo {author} {\bibfnamefont {L.~J.}\ \bibnamefont {Papenfort}}, \bibinfo {author} {\bibfnamefont {S.~D.}\ \bibnamefont {Tootle}}, \bibinfo {author} {\bibfnamefont {P.}~\bibnamefont {Grandcl\'ement}}, \bibinfo {author} {\bibfnamefont {E.~R.}\ \bibnamefont {Most}},\ and\ \bibinfo {author} {\bibfnamefont {L.}~\bibnamefont {Rezzolla}},\ }\bibfield  {title} {\bibinfo {title} {{New public code for initial data of unequal-mass, spinning compact-object binaries}},\ }\href {https://doi.org/10.1103/PhysRevD.104.024057} {\bibfield  {journal} {\bibinfo  {journal} {Phys. Rev. D}\ }\textbf {\bibinfo {volume} {104}},\ \bibinfo {pages} {024057} (\bibinfo {year} {2021})},\ \Eprint {https://arxiv.org/abs/2103.09911} {arXiv:2103.09911 [gr-qc]} \BibitemShut {NoStop}%
\bibitem [{\citenamefont {Most}\ and\ \citenamefont {Raithel}(2021)}]{Most:2021ktk}%
  \BibitemOpen
  \bibfield  {author} {\bibinfo {author} {\bibfnamefont {E.~R.}\ \bibnamefont {Most}}\ and\ \bibinfo {author} {\bibfnamefont {C.~A.}\ \bibnamefont {Raithel}},\ }\bibfield  {title} {\bibinfo {title} {{Impact of the nuclear symmetry energy on the post-merger phase of a binary neutron star coalescence}},\ }\href {https://doi.org/10.1103/PhysRevD.104.124012} {\bibfield  {journal} {\bibinfo  {journal} {Phys. Rev. D}\ }\textbf {\bibinfo {volume} {104}},\ \bibinfo {pages} {124012} (\bibinfo {year} {2021})},\ \Eprint {https://arxiv.org/abs/2107.06804} {arXiv:2107.06804 [astro-ph.HE]} \BibitemShut {NoStop}%
\end{thebibliography}%

\end{document}